\documentclass[prd,showpacs,showkeys,preprintnumbers,notitlepage,nofootinbib]{revtex4}
\usepackage[latin9]{inputenc}
\usepackage[letterpaper]{geometry}
\usepackage{amsmath}
\usepackage{amssymb}
\usepackage{graphicx}
\usepackage{esint}
\usepackage[unicode=true,pdfusetitle,
 bookmarks=true,bookmarksnumbered=false,bookmarksopen=false,
 breaklinks=false,pdfborder={0 0 1},backref=false,colorlinks=false]
 {hyperref}
\usepackage{breakurl}

\makeatletter

\providecommand{\tabularnewline}{\\}
\newcommand{\lyxdot}{.}

\@ifundefined{textcolor}{}
{%
 \definecolor{BLACK}{gray}{0}
 \definecolor{WHITE}{gray}{1}
 \definecolor{RED}{rgb}{1,0,0}
 \definecolor{GREEN}{rgb}{0,1,0}
 \definecolor{BLUE}{rgb}{0,0,1}
 \definecolor{CYAN}{cmyk}{1,0,0,0}
 \definecolor{MAGENTA}{cmyk}{0,1,0,0}
 \definecolor{YELLOW}{cmyk}{0,0,1,0}
}

\providecommand{\tabularnewline}{\\}

\def\GeV{\,\mbox{GeV}}

\@ifundefined{textcolor}{}{%
 \definecolor{BLACK}{gray}{0}
 \definecolor{WHITE}{gray}{1}
 \definecolor{RED}{rgb}{1,0,0}
 \definecolor{GREEN}{rgb}{0,1,0}
 \definecolor{BLUE}{rgb}{0,0,1}
 \definecolor{CYAN}{cmyk}{1,0,0,0}
 \definecolor{MAGENTA}{cmyk}{0,1,0,0}
 \definecolor{YELLOW}{cmyk}{0,0,1,0}
}

\providecommand{\tabularnewline}{\\}

\makeatother

\begin{document}

\title{Bethe-Heitler type radiative corrections\\
 to deeply virtual neutrinoproduction of mesons}

\author{B.~Z.~Kopeliovich, Iván~Schmidt and M.~Siddikov}

\address{Departamento de F\'{i}sica, y Centro Cient\'{i}fico - Tecnológico
de Valpara\'{i}so, Universidad Técnica Federico Santa Mar\'{i}a,
Casilla 110-V, Valpara\'{i}so, Chile}

\preprint{USM-TH-311}
\begin{abstract}
We study the electromagnetic Bethe-Heitler type contribution to neutrino-induced
deeply virtual meson production ($\nu$DVMP). Such $\mathcal{O}(\alpha_{em})$-corrections
decrease with $Q^{2}$ in the Bjorken regime less steeply than the
standard $\nu$DVMP handbag contribution. Therefore, they are relatively
enhanced at high $Q^{2}$. The Bethe-Heitler terms give rise to an
angular correlation between the lepton and hadron scattering planes
with harmonics sensitive to the real and imaginary parts of the DVMP
amplitude. These corrections constitute a few percent effect in the
kinematics of the forthcoming \textsc{Minerva} experiment at Fermilab
and should be taken into account in precision tests of GPD parametrizations.
For virtualities $Q^{2}\sim100$~GeV$^{2}$ these corrections become
on a par with DVMP handbag contributions. A computational code, which
can be used for the evaluation of these corrections employing various
GPD models is provided. 
\end{abstract}

\pacs{13.15.+g,13.85.-t}

\keywords{Single pion production, generalized parton distributions, Bethe-Heitler
contributions}

\maketitle

\section{Introduction}

Generalized parton distributions (GPD) allow evaluation of cross-sections
for a wide class of processes, where the collinear factorization is
applicable~\cite{Ji:1998xh,Collins:1998be}. The main source of experimental
information on GPDs has been so far the electron(positron)-proton
measurements performed at JLAB and HERA, in particular deeply virtual
Compton scattering (DVCS) and deeply virtual meson production (DVMP)~\cite{Mueller:1998fv,Ji:1996nm,Ji:1998pc,Radyushkin:1996nd,Radyushkin:1997ki,Radyushkin:2000uy,Ji:1998xh,Collins:1998be,Collins:1996fb,Brodsky:1994kf,Goeke:2001tz,Diehl:2000xz,Belitsky:2001ns,Diehl:2003ny,Belitsky:2005qn,Kubarovsky:2011zz}.
The 12~GeV upgrade at Jefferson lab will open new opportunities for
further improvement of our knowledge of the GPDs~\cite{Kubarovsky:2011zz}.

However, the practical realization of this program suffers from large
uncertainties. For instance, the results at moderately high $Q^{2}$
can be affected by poorly known higher-twist components of GPDs and
distribution amplitudes (DA) of the produced mesons~\cite{Ahmad:2008hp,Goloskokov:2009ia,Goloskokov:2011rd,Goldstein:2012az}.

Neutrino experiments provide a powerful tool for consistency checks
for the extraction of GPD from JLAB data, especially of their flavor
structure. The study of various processes in the Bjorken regime may
be done with the high-intensity \textsc{NuMI} beam at Fermilab, which
will switch soon to the so-called middle-energy (ME) regime with an
average neutrino energy of about $6$~GeV, and potentially may reach
energies up to 20 GeV, without essential loss of luminosity. In this
setup the \textsc{Minerva} experiment\textsc{~\cite{Drakoulakos:2004gn}}
should be able to probe the quark flavor structure of the targets.
Even higher luminosities in multi-GeV regime can be achieved at the
planned Muon Collider/Neutrino Factory~\cite{Gallardo:1996aa,Ankenbrandt:1999as,Alsharoa:2002wu}.

Certain information on the GPD flavor structure can be extracted from
comparison of analogous processes in neutrino- and electro-induced
processes employing the difference of flavor structures of electromagnetic
and weak neutral currents. An example is the weak DVCS~\cite{Psaker:2006gj},
which alone, however, is not sufficient to constrain the flavor structure.

Recently we discussed the possibility of GPD extraction from deeply
virtual neutrino-production of the pseudo-Goldstone mesons ($\pi,\, K,\,\eta$)~\cite{Kopeliovich:2012dr}.
The $\nu$DVMP measurements with neutrino and antineutrino beams are
complementary to the electromagnetic DVMP. The octet of pseudo-Goldstone
bosons, originating from the chiral symmetry breaking, acts in the
axial current as a natural probe for the flavor content. Due to the
$V-A$ structure of the charged current, in $\nu$DVMP one can access
simultaneously the unpolarized GPDs, $H$ and $E$, and the helicity
flip GPDs, $\tilde{H}$ and $\tilde{E}$. We expect the contributions
of the GPDs $H_{T},\, E_{T},\,\tilde{H}_{T}$ and $\tilde{E}_{T}$,
which are controlled by the poorly known twist-3 pion DA $\phi_{p}$,
to be negligible. Besides, important information on the flavor structure
can be obtained by studying the transitional GPDs in the processes
with nucleon to hyperon transitions. As was discussed in~\cite{Frankfurt:1999fp},
assuming $SU(3)$ flavor symmetry one can relate these GPDs to the
ordinary diagonal ones in the proton.

In this paper we study the Bethe-Heitler (BH) type radiative corrections
to the diffractive neutrino-production of charged pseudo-Golstone
mesons with the target remained intact, related to meson emission
from lepton line with subsequent electromagnetic interaction with
the target. Although such processes are formally suppressed by $\alpha_{em}$,
at large $Q^{2}$ they fall off less steeply than the DVMP cross-section.
While at virtualities $Q^{2}\lesssim10$ GeV$^{2}$, relevant for
modern neutrino experiments, this is a few percent correction, already
at $Q^{2}\sim100$~GeV$^{2}$ it becomes comparable with the DVMP
cross section. Such corrections are relevant only in case of the $\nu$DVMP:
for the electron-induced DVMP $ep\to ep\, M$ they are suppressed
by factor $\sim\left(G_{F}Q^{2}\right)^{2}$, where $G_{F}$ is the
Fermi coupling, and are negligible unless we consider extremely high
$Q^{2}\approx M_{W}^{2}$. Existence of such diagrams opens a possibility
to probe separately the real and imaginary parts of the DVMP amplitude
(not only the total cross-sections), in close analogy to DVCS studies~\cite{Belitsky:2001ns,Belitsky:2002tf,Belitsky:2003fj}.

The paper is organized as follows. In Section~\ref{sec:DVMP_Xsec}
we present the results for the DVMP and BH contributions to the experimentally
measurable total $\nu$DVMP cross-section (technical details of evaluation
may be found in Appendix~\ref{sec:DVMP_Xsec-App}). Also, at the
end of Section~\ref{sec:DVMP_Xsec} we construct two asymmetries
which have particular sensitivity to the real and imaginary parts
of the DVMP amplitude. In Section~\ref{sec:Parametrizations}, for
the sake of completeness we discuss the features of the GPD parametrization
used in calculations. In Section~\ref{sec:Results} we present the
numerical results and make conclusions.

\section{Cross-section of the $\nu$DVMP and BH processes}

\label{sec:DVMP_Xsec}

\begin{figure}[htp]
\includegraphics[scale=0.35]{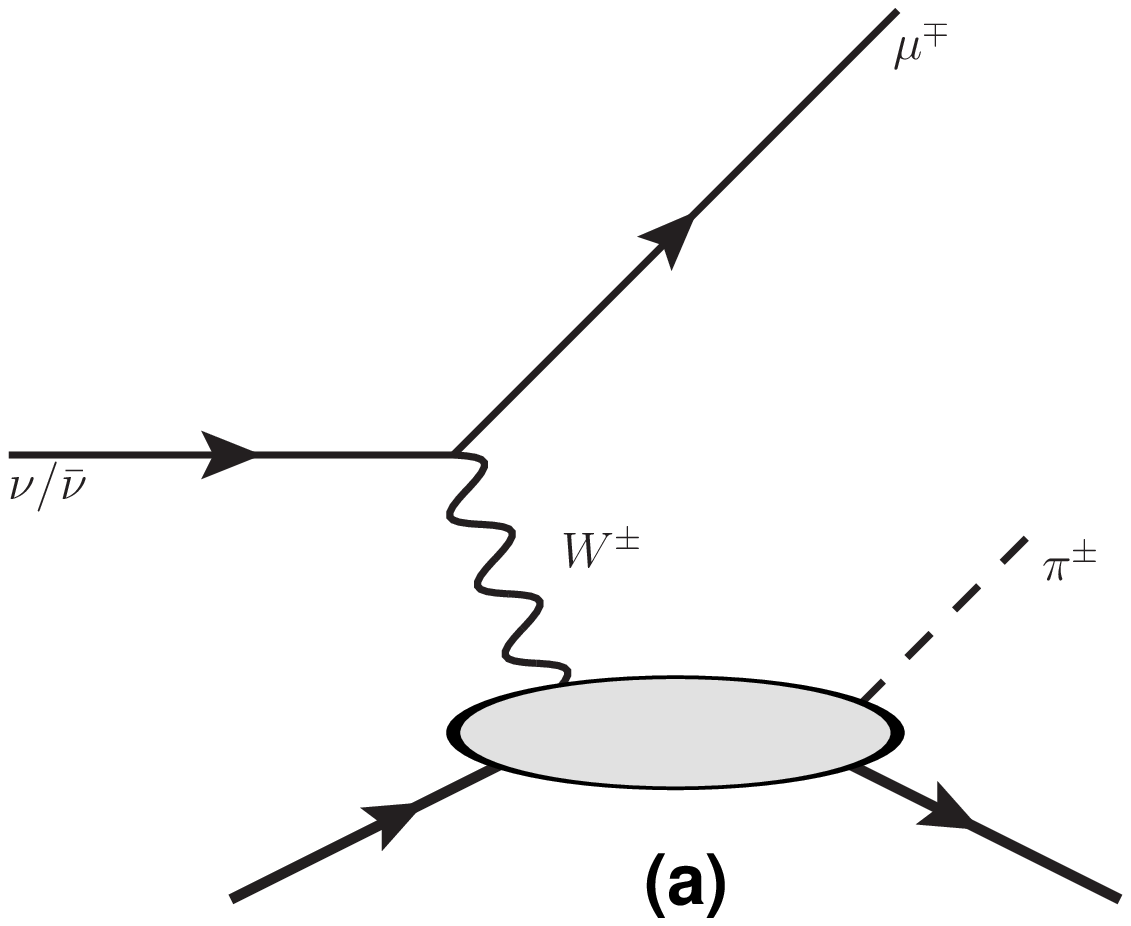}\includegraphics[scale=0.35]{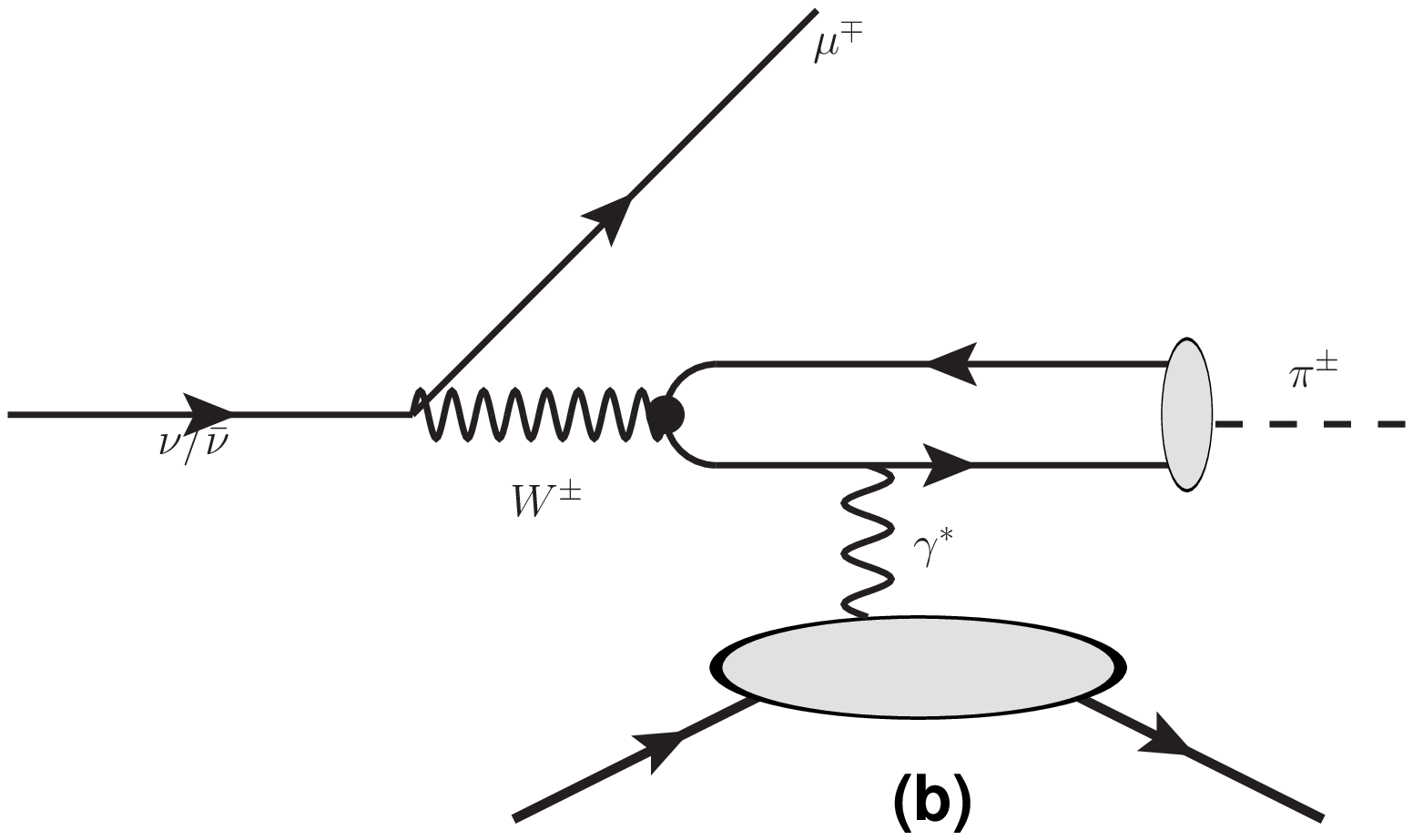}\includegraphics[scale=0.35]{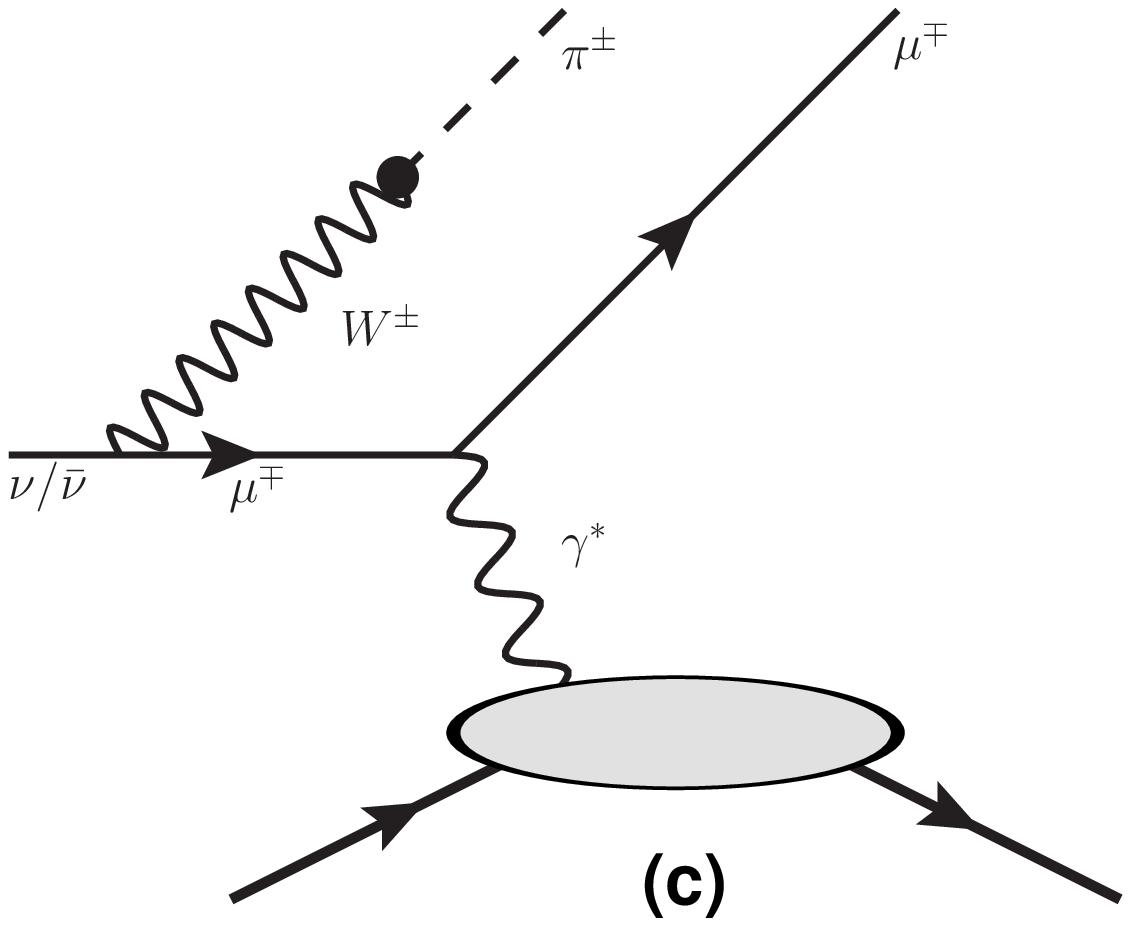}\caption{\label{fig:DVMPLT}Diagrams contributing to the neutrinoproduction
of mesons. (a) DVMP process (b,c) BH contributions.}
\end{figure}

Exclusive neutrino-production of pions is presented by the diagram
(a) in the Figure~\ref{fig:DVMPLT}. Production of pions by the vector
current was studied in~\cite{Vanderhaeghen:1998uc,Mankiewicz:1998kg},
and was recently extended to neutrino interactions in~\cite{Kopeliovich:2012dr}.

As was shown in~\cite{Vanderhaeghen:1998uc,Mankiewicz:1998kg}, at
large $Q^{2}$ where the collinear factorization is applicable, the
amplitude of this process is suppressed due to hard gluon exchange
in the coefficient function, and is small. This raises a question,
how important could be the $\mathcal{O}\left(\alpha_{em}\right)$
corrections ? While a systematic study of radiative corrections is
beyond the scope of the present paper, we would like to consider in
detail the contributions which decrease with $Q^{2}$ less rapidly
than the diagram (a). In the leading order in $\alpha_{em}$ and $\alpha_{W}$
there are two such diagrams (b,c) shown in the Figure~\ref{fig:DVMPLT}.
These diagrams are enhanced by a factor $\sim Q^{2}/t$, where $-Q^{2}\equiv q^{2}$
and $t$ are the lepton and proton 4-momentum transfers squared, respectively.
This factor is large in the Bjorken regime of $t\ll Q^{2}$, and despite
the formal suppression by $\alpha_{em}$, the diagrams (b,c) are numerically
comparable to the diagram (a). Notice that such corrections are sizable
only in the case of neutrino-production of mesons. In case of electroproduction
the corresponding corrections are suppressed by $\sim G_{F}^{2}Q^{4}$,
where $G_{F}$ is the Fermi constant, and are negligible, unless we
go to extremely large $Q^{2}\sim M_{W}^{2}$ . In what follows we
evaluate the contribution of each diagram in Figure~\ref{fig:DVMPLT}.

The cross-section of pion production can be presented as a sum of
contributions of DVMP (diagram (a)), BH mechanism (diagrams (b,c))
and their interference, 
\begin{equation}
\frac{d^{4}\sigma}{dt\, d\ln x_{Bj}\, dQ^{2}d\phi}=\frac{d^{4}\sigma^{(DVMP)}}{dt\, d\ln x_{Bj}\, dQ^{2}d\phi}+\frac{d^{4}\sigma^{(BH)}}{dt\, d\ln x_{Bj}\, dQ^{2}d\phi}+\frac{d^{4}\sigma^{(int)}}{dt\, d\ln x_{Bj}\, dQ^{2}d\phi},\label{eq:XSec_sum}
\end{equation}
where $x_{Bj}=Q^{2}/(2P\cdot q)$, and $\phi$ is the angle between
the lepton and hadron scattering planes.

Evaluation of the diagram (a) is straightforward and yields~\cite{Kopeliovich:2012dr}
\begin{eqnarray}
\frac{d^{4}\sigma^{(DVMP)}}{dt\, d\ln x_{Bj}\, dQ^{2}d\phi} & = & \frac{G_{F}^{2}f_{M}^{2}x_{Bj}^{2}\left(1-y-\frac{m_{N}^{2}x_{Bj}^{2}y^{2}}{Q^{2}}\right)}{64\,\pi^{4}Q^{2}\left(1+Q^{2}/M_{W}^{2}\right)^{2}\left(1+\frac{4m_{N}^{2}x_{Bj}^{2}}{Q^{2}}\right)^{3/2}}\left|T_{M}\right|^{2},\label{eq:XSec_DVMP}
\end{eqnarray}
where $y$ is the fractional loss of lepton energy defined as $y=P\cdot q/P\cdot k=\nu/E_{\nu}$.
Notice that the DVMP cross section turns out to be independent of
the angle $\phi$ between lepton and hadron planes. This happens because
the momentum $q$ does not have transverse components in the Bjorken
reference frame.

For unpolarized target, the matrix element squared $\left|T_{M}\right|^{2}$
in Eqn.~(\ref{eq:XSec_DVMP}) can be simplified as, 
\begin{eqnarray}
\left|T_{M}\right|_{unp}^{2} & = & \frac{64\pi^{2}}{81}\frac{\alpha_{s}^{2}}{Q^{2}(2-x_{Bj})^{2}}\phi_{-1}^{2}4\left[4\left(1-x_{Bj}\right)\left(\mathcal{H}_{M}\mathcal{H}_{M}^{*}+\tilde{\mathcal{H}}_{M}\tilde{\mathcal{H}}_{M}^{*}\right)-\frac{x_{Bj}^{2}t}{4m_{N}^{2}}\tilde{\mathcal{E}}_{M}\tilde{\mathcal{E}}_{M}^{*}\right.\label{eq:T2_unp}\\
 & - & \left.x_{Bj}^{2}\left(\mathcal{H}_{M}\mathcal{E}_{M}^{*}+\mathcal{E}_{M}\mathcal{H}_{M}^{*}+\tilde{\mathcal{H}}_{M}\tilde{\mathcal{E}}_{M}^{*}+\tilde{\mathcal{E}}_{M}\tilde{\mathcal{H}}_{M}^{*}\right)-\left(x_{Bj}^{2}+\left(2-x_{Bj}\right)^{2}\frac{t}{4m_{N}^{2}}\right)\mathcal{E}_{M}\mathcal{E}_{M}^{*}\right],\nonumber 
\end{eqnarray}
where we introduced a shorthand notation, 
\begin{equation}
\phi_{-1}=\int_{0}^{1}dz\frac{\phi_{M}(z)}{z}=\frac{1}{2}\int_{0}^{1}dz\frac{\phi_{M}(z)}{z\bar{z}},\label{eq:MinusFirstMoment}
\end{equation}
and the script letters $\mathcal{H},\,\mathcal{E},\,\tilde{\mathcal{H}},\,\tilde{\mathcal{E}}$
signify convolution of the GPDs $H,\, E,\,\tilde{H},\,\tilde{E}$
with corresponding coefficient functions given in Table~\ref{tab:DVMP_amps}.
\begin{table}[h]
\caption{\label{tab:DVMP_amps}List of the DVMP amplitudes $\mathcal{H}_{M},\,\mathcal{E}_{M},\,\tilde{\mathcal{H}}_{M},\,\tilde{\mathcal{E}}_{M}$
for different final states. For a neutron target, in the r.h.s. we
flipped $H_{u/n}\to H_{d/p}$,~$H_{d/n}\to H_{u/p}$, so all the
GPDs are given for a proton target. To get $\mathcal{E}$, $\tilde{\mathcal{H}}$,
$\tilde{\mathcal{E}}$ one should replace $H$ with $E$, $\tilde{H}$
and $\tilde{E}$ respectively. $V_{ij}$ are the CKM matrix elements.
$c_{\pm}$ is a shorthand notation $c_{\pm}(x,\xi)=1/(x\pm\xi\mp i0)$
for the leading order coefficient function. The NLO corrections to
the coefficient functions may be found in~\cite{Ivanov:2004zv,Diehl:2007hd}.
For the sake of brevity we do not show the arguments $(x,\xi,t,Q)$
for all GPDs and omitted the integral over the quark light-cone fraction
$\int dx$ everywhere. }

\global\long\def\arraystretch{1.5}
\begin{tabular}{|c|c|c|c|c|c|c|}
\hline 
Process  & type  & $\mathcal{H}_{M}$  &  & Process  & type  & $\mathcal{H}_{M}$\tabularnewline
\hline 
$\nu\, p\to\mu^{-}\pi^{+}p$  & CC  & $V_{ud}\left(H_{d}c_{-}+H_{u}c_{+}\right)$  &  & $\nu\, n\to\mu^{-}\pi^{+}n$  & CC  & $V_{ud}\left(H_{u}c_{-}+H_{d}c_{+}\right)$\tabularnewline
\hline 
$\bar{\nu}\, p\to\mu^{+}\pi^{-}p$  & CC  & $V_{ud}\left(H_{u}c_{-}+H_{d}c_{+}\right)$  &  & $\bar{\nu}\, n\to\mu^{+}\pi^{-}n$  & CC  & $V_{ud}\left(H_{d}c_{-}+H_{u}c_{+}\right)$\tabularnewline
\cline{1-3} \cline{5-7} 
\multicolumn{1}{c}{} & \multicolumn{1}{c}{} & \multicolumn{1}{c}{} & \multicolumn{1}{c}{} & \multicolumn{1}{c}{} & \multicolumn{1}{c}{} & \multicolumn{1}{c}{}\tabularnewline
\hline 
$\nu\, p\to\mu^{-}K^{+}p$  & CC  & $V_{us}\left(c_{+}H_{u}+c_{-}H_{s}\right)$  &  & $\nu\, n\to\mu^{-}K^{+}n$  & CC  & $V_{us}\left(c_{+}H_{d}+c_{-}H_{s}\right)$\tabularnewline
\hline 
$\bar{\nu}\, p\to\mu^{+}K^{-}p$  & CC  & $V_{us}\left(H_{u}c_{-}+H_{s}c_{+}\right)$  &  & $\bar{\nu}\, n\to\mu^{+}K^{-}n$  & CC  & $V_{us}\left(H_{d}c_{-}+H_{s}c_{+}\right)$\tabularnewline
\hline 
\end{tabular}
\end{table}

Comparing different elements in Table~\ref{tab:DVMP_amps}, one gets
relations, 
\begin{align}
d\sigma_{\bar{\nu}\, p\to\mu^{+}\pi^{-}p}^{DVMP} & =d\sigma_{\nu\, n\to\mu^{-}\pi^{+}n}^{DVMP},\label{eq:isospin_DVMP_1}\\
d\sigma_{\nu\, p\to\mu^{-}\pi^{+}p}^{DVMP} & =d\sigma_{\bar{\nu}\, n\to\mu^{+}\pi^{-}n}^{DVMP},\label{eq:isospin_DVMP_2}
\end{align}
which are just a manifestation of the isospin symmetry. As shown below,
these relations are broken by the BH corrections.

In the leading order in $Q^{2}$ both BH diagrams, Fig.~\ref{fig:DVMPLT}
(b) and (c), acquire dominant contribution from longitudinally polarized
photons. However, as discussed below, certain angular harmonics are
suppressed by $\sim\Delta_{\perp}/Q$ and get similar contributions
from transverse and longitudinal photons. For this reason, for the
form factors we include both the longitudinal and transverse components
and evaluate the BH diagrams exactly, and only after that we make
an expansion in $Q^{2}$. The diagram (b) contains the matrix element

\begin{equation}
\mathcal{A}_{\mu\nu}^{ab}\left(q,\Delta\right)=\frac{1}{f_{\pi}}\int d^{4}x\, e^{-iq\cdot x}\left\langle 0\left|\left(V_{\mu}^{a}(x)-A_{\mu}^{a}(x)\right)J_{\nu}^{em}(0)\right|\pi^{b}\left(q-\Delta\right)\right\rangle ,\label{eq:A_munu_def}
\end{equation}
where $V_{\mu}^{a}(x)$ and $A_{\mu}^{a}(x)$ are the vector and axial
vector isovector currents. The correlator~(\ref{eq:A_munu_def})
 can be evaluated in pQCD in the collinear approximation, because
$Q^{2}$ is large, and we assume that the dominant contribution comes
from the leading twist-2 pion DA. Notice that the amplitude $\mathcal{A}_{\mu\nu}$
cannot be interpreted as the pion form factor because (i) the virtuality
of $W$ is large, (ii) the insertion of the pion state between $A_{\mu}^{a}$
and $J_{\nu}^{em}$ leads to $\mathcal{A}_{\mu\nu}^{ab}\sim q_{\mu}$,
which gives zero acting on the transverse on-shell lepton current.
The details of the calculations are moved to Appendix~\ref{sec:DVMP_Xsec-App},
and we present here the final result, which reads,

\begin{equation}
\frac{d^{4}\sigma^{(BH)}}{dt\, d\ln x_{Bj}\, dQ^{2}d\phi}=\frac{f_{\pi}^{2}G_{F}^{2}\alpha_{em}^{2}\sum_{n=0}^{2}\mathcal{C}_{n}^{BH}\cos(n\phi)}{16\,\pi^{2}t^{2}\left(1+\frac{4m_{N}^{2}x_{B}^{2}}{Q^{2}}\right)^{5/2}},\label{eq:XSec_BH}
\end{equation}
where 
\begin{align}
\mathcal{C}_{0}^{BH} & =\mathcal{C}_{2}^{BH}+\frac{m_{N}^{2}}{9\, Q^{2}}\left[4\left(\left(2y^{2}+y-1\right)\left(\phi_{-1}-1\right)x_{B}^{3}\right.\right.\\
 & -\left(\left(4\left(\phi_{-1}-1\right)^{2}+\frac{t}{2m_{N}^{2}}\left(4\phi_{-1}^{2}-8\phi_{-1}+5\right)\right)y^{2}-4\left(\left(\phi_{-1}-1\right)^{2}+\frac{t}{4m_{N}^{2}}\left(4\phi_{-1}^{2}-13\phi_{-1}+10\right)\right)y\right.\nonumber \\
 & +\left.\frac{5t}{2m_{N}^{2}}\left(\phi_{-1}-2\right)^{2}+\left(\phi_{-1}-1\right)^{2}\right)x_{B}^{2}+\left(2y-1\right)\frac{t}{m_{N}^{2}}\left(\phi_{-1}-1\right)\left(-\phi_{-1}+y\left(2\phi_{-1}-1\right)+2\right)x_{B}\nonumber \\
 & -\left.4\left(1-2y\right)^{2}\frac{t}{4m_{N}^{2}}\left(\phi_{-1}-1\right)^{2}\right)F_{1}^{2}(t)
+2 F_{1}(t)F_{2}(t)x_{B}^{2}\left(x_{B}^{2}\left(y+1\right)^{2}-\frac{x_{B}t}{m_{N}^{2}}\left(y+1\right)^{2}\right.\nonumber \\
 & \left.-\frac{t}{m_{N}^{2}}\left(\left(8\phi_{-1}^{2}-24\phi_{-1}+17\right)y^{2}-2\left(8\phi_{-1}^{2}-24\phi_{-1}+19\right)y+10\phi_{-1}^{2}-36\phi_{-1}+35\right)\right)\nonumber \\
 & +F_{2}^{2}\left(\left(y+1\right)^{2}\left(\frac{t}{4m_{N}^{2}}+1\right)x_{B}^{4}-\left(y+1\right)\frac{t}{m_{N}^{2}}\left(\frac{t}{4m_{N}^{2}}-\phi_{-1}+y\left(\frac{t}{4m_{N}^{2}}+2\phi_{-1}-1\right)+2\right)x_{B}^{3}\right.\nonumber \\
 & +\frac{t\, x_{B}^{2}}{m_{N}^{2}}\left(\left(-4\phi_{-1}^{2}+16\phi_{-1}+\frac{t}{4m_{N}^{2}}\left(8\phi_{-1}-7\right)-13\right)y^{2}\right.\nonumber \\
 & \left.+2\left(6\phi_{-1}^{2}-20\phi_{-1}+\frac{t}{4m_{N}^{2}}\left(2\phi_{-1}-1\right)+17\right)y-9\phi_{-1}^{2}+\frac{t}{4m_{N}^{2}}\left(5-4\phi_{-1}\right)+34\phi_{-1}-34\right)\nonumber \\
 & \left.\left.-\left(2y-1\right)\left(\frac{t}{m_{N}^{2}}\right)^{2}\left(\phi_{-1}-1\right)\left(-\phi_{-1}+y\left(2\phi_{-1}-1\right)+2\right)x_{B}+\left(1-2y\right)^{2}\left(\frac{t}{4m_{N}^{2}}\right)^{2}\left(\phi_{-1}-1\right)^{2}\right)\right]\nonumber \\
 & +\mathcal{O}\left(\frac{m_{N}^{4}}{Q^{4}},\frac{t^{2}}{Q^{4}}\right),\nonumber 
\end{align}

\begin{align}
\mathcal{C}_{1}^{BH} & =\frac{K\, m_{N}^{2}}{9Q^{2}}\left[4\left(3\left(-4y+3\left(y-2\right)\phi_{-1}+9\right)x_{B}^{3}\right.\right.\\
 & -2\left(\phi_{-1}-1\right)\left(-\left(\frac{5t}{2m_{N}^{2}}+9\right)\phi_{-1}+2y\left(\frac{t\phi_{-1}}{m_{N}^{2}}-\frac{3t}{4m_{N}^{2}}+3\phi_{-1}-6\right)+18\right)x_{B}^{2}\nonumber \\
 & +\left.\frac{3t}{m_{N}^{2}}\left(\phi_{-1}-1\right)\left(-6\phi_{-1}+y\left(4\phi_{-1}-3\right)+3\right)x_{B}-\left(2y-3\right)\frac{6\, t}{m_{N}^{2}}\left(\phi_{-1}-1\right)^{2}\right)F_{1}^{2}(t)\nonumber \\
 & +4F_{1}(t)F_{2}(t)x_{B}^{2}\left(3\left(y-3\right)x_{B}^{2}-12\left(y-3\right)\frac{t}{4m_{N}^{2}}x_{B}-8\frac{t}{4m_{N}^{2}}\left(4y\left(\phi_{-1}-3\right)-5\phi_{-1}+18\right)\left(\phi_{-1}-1\right)\right)\nonumber \\
 & -F_{2}^{2}(t)\left(-6\left(y-3\right)x_{B}^{4}\right.\nonumber \\
 & +\frac{t}{4m_{N}^{2}}\left(-6\left(y-3\right)x_{B}^{2}+12\left(-2y+3\left(y-2\right)\phi_{-1}+3\right)x_{B}+8\left(2y\left(\phi_{-1}-6\right)-\phi_{-1}+18\right)\left(\phi_{-1}-1\right)\right)x_{B}^{2}\nonumber \\
 & \left.\left.+24\left(\frac{t}{4m_{N}^{2}}\right)^{2}\left(x_{B}\left(x_{B}-2\phi_{-1}+2\right)+2\left(\phi_{-1}-1\right)\right)\left(x_{B}\left(y-3\right)-2\left(2y-3\right)\left(\phi_{-1}-1\right)\right)\right)\right]\nonumber \\
 & +\mathcal{O}\left(\frac{m_{N}^{4}}{Q^{4}},\frac{t^{2}}{Q^{4}}\right)\nonumber 
\end{align}
\begin{align}
\mathcal{C}_{2}^{BH} & =-\frac{4\, K^{2}}{9}\left[\left(5x_{B}-4\phi_{-1}+4\right)\left(\phi_{-1}-1\right)F_{1}^{2}(t)+2x_{B}^{2}F_{1}(t)F_{2}(t)\right.\\
 & +\left.\left(\left(1+\frac{t}{4m_{N}^{2}}\right)x_{B}^{2}-\frac{5\, t\, x_{B}}{4m_{N}^{2}}\left(\phi_{-1}-1\right)+\frac{t}{m_{N}^{2}}\left(\phi_{-1}-1\right)^{2}\right)F_{2}^{2}(t)\right]\nonumber \\
 & +\mathcal{O}\left(\frac{m_{N}^{2}}{Q^{2}},\,\frac{t}{Q^{2}}\right).\nonumber 
\end{align}

Here, following Refs.~\cite{Belitsky:2001ns,Belitsky:2002tf,Belitsky:2003fj}
we introduced the notations,

\begin{align}
K^{2} & =\frac{\Delta_{\perp}^{2}}{Q^{2}}\left(1-y-\frac{y^{2}\epsilon^{2}}{4}\right)\\
 & =-\frac{t}{Q^{2}}\left(1-x_{B}\right)\left(1-y-\frac{\epsilon^{2}y^{2}}{4}\right)\left(1-\frac{t_{min}}{t}\right)\left\{ \sqrt{1+\epsilon^{2}}+\frac{4x_{B}\left(1-x_{B}\right)+\epsilon^{2}}{4\left(1-x_{B}\right)}\frac{t-t_{min}}{Q^{2}}\right\} ,\nonumber \\
\epsilon^{2} & =\frac{4m_{N}^{2}x_{B}^{2}}{Q^{2}},\quad t_{min}=-\frac{m_{N}^{2}x_{B}^{2}}{1-x_{B}}+\mathcal{O}\left(\frac{m_{N}^{2}}{Q^{2}},\frac{t}{Q^{2}}\right).
\end{align}

Notice that in the leading order in $Q^{2}$ the contributions of
diagrams (b) and (c) to the terms $\mathcal{C}_{0}^{BH}$ and $\mathcal{C}_{1}^{BH}$
exactly cancel each other, so that the coefficients $\mathcal{C}_{0}^{BH}$
and $\mathcal{C}_{1}^{BH}$ acquire an extra suppression factor $m_{N}^{2}/Q^{2}$.
This can be understood as a screening of the opposite charges of the
pion and muon, when both of them move in the forward direction in
the limit of massless leptons. As we can see, the BH cross-section
is symmetric under the $\phi\to-\phi$ transformation. For asymptotically
large $Q^{2}$, the $\mathcal{C}_{1}^{BH}$ harmonic is suppressed
by $\Delta_{\perp}/Q$, whereas $\mathcal{C}_{0}^{BH}\sim\mathcal{C}_{2}^{BH}$,
so the distribution is symmetric relative to the replacement $\phi\to\pi-\phi$.
Note also that $K\sim\Delta_{\perp}^{2}$ and vanishes when $\Delta_{\perp}\to0$.

The interference term in~(\ref{eq:XSec_sum}) has a form (see details
in Appendix~\ref{sec:DVMP_Xsec-App})

\begin{align}
\frac{d^{4}\sigma^{(int)}}{dt\, d\ln x_{Bj}\, dQ^{2}d\phi} & =\frac{f_{\pi}^{2}G_{F}^{2}\, x_{B}\alpha_{em}\alpha_{S}\phi_{-1}\left(\mathcal{C}_{0}^{int}+\mathcal{C}_{1}^{int}\cos\phi+\mathcal{S}_{1}^{int}\sin\phi\right)}{36\,\pi^{2}t\, Q^{2}\left(1-\frac{x_{B}}{2}\right)\left(1+\frac{4m_{N}^{2}x_{B}^{2}}{Q^{2}}\right)^{5/2}},\label{eq:XSec_Int}
\end{align}
where 
\begin{align}
\mathcal{C}_{0}^{int} & =-\frac{m_{N}^{2}(1-y)}{Q^{2}}\left(\left(-4\left(1-x_{B}\right)\Re e\mathcal{H}+x_{B}^{2}\,\Re e\mathcal{E}\right)F_{1}+\Re e\mathcal{E}\, F_{2}\frac{t}{4m_{N}^{2}}(x_{B}-2)^{2}+\left(\Re e\mathcal{H}+\Re e\mathcal{E}\right)F_{2}x_{B}^{2}\right)\label{eq:Int_C0}\\
 & \times\left(\left(2\phi_{-1}-3\right)x_{B}^{2}-\frac{t}{m_{N}^{2}}\left(\phi_{-1}-1\right)\left(1+x_{B}\right)\right)+\mathcal{O}\left(\frac{m_{N}^{2}}{Q^{2}},\frac{t}{Q^{2}}\right),\nonumber \\
\mathcal{C}_{1}^{int} & =\frac{K}{3}\,\left[F_{1}\left(2\,\mathcal{R}e\,\mathcal{E}\left(y-3\right)x_{B}^{2}+\mathcal{R}e\,\mathcal{H}\left(4\left(x_{B}-2\right)\left(2y-3\right)\phi_{-1}-4\left(\left(x_{B}-4\right)y+6\right)\right)\right)\right.\label{eq:Int_C1}\\
 & +F_{2}\left(2\,\mathcal{R}e\mathcal{H}\left(y-3\right)x_{B}^{2}+\mathcal{R}e\,\mathcal{E}\left(2x_{B}^{2}\left(y-3\right)-\left(x_{B}-2\right)\frac{t}{4m_{N}^{2}}\left(4\left(2y-3\right)\left(\phi_{-1}-1\right)-2x_{B}\left(y-3\right)\right)\right)\right)\nonumber \\
 & +\left.\mathcal{O}\left(\frac{m_{N}^{2}}{Q^{2}},\frac{t}{Q^{2}}\right)\right]\nonumber \\
\mathcal{S}_{1}^{int} & =\frac{K\left(2-x_{B}\right)}{6}\,\left[F_{1}\left(-2\,\mathcal{I}m\,\mathcal{E}\left(y+1\right)x_{B}^{2}-2\,\mathcal{I}m\,\mathcal{H}\left(x_{B}\left(4\phi_{-1}y-2y-2\phi_{-1}+4\right)-4\left(2y-1\right)\left(\phi_{-1}-1\right)\right)\right)\right.\label{eq:Int_S1}\\
 & +F_{2}\left(-2\,\mathcal{I}m\,\mathcal{H}\left(y+1\right)x_{B}^{2}\right.\nonumber \\
 & \left.-2\,\mathcal{I}m\,\mathcal{E}\left(\left(y+1\right)\left(1+\frac{t}{4m_{N}^{2}}\right)x_{B}^{2}+\frac{t}{2m_{N}^{2}}\left(-2\phi_{-1}y+y+\phi_{-1}-2\right)x_{B}+\left(2y-1\right)\frac{t}{m_{N}^{2}}\left(\phi_{-1}-1\right)\right)\right)\nonumber \\
 & +\left.\mathcal{O}\left(\frac{m_{N}^{2}}{Q^{2}},\frac{t}{Q^{2}}\right)\right],\nonumber 
\end{align}

As one can see from~(\ref{eq:XSec_Int}), the angular dependence
of the interference term has a $\sin\phi$ term which is absent both
in the BH and DVMP taken alone. This term stems from the interference
of the vector and axial vector current in lepton part of the diagram.
It has the same sign for neutrino and antineutrino beams (in contrast,
all the other terms in~(\ref{eq:XSec_Int}) change sign). Such terms
are absent in case of DVMP and BH since the interference is asymmetric
w.r.t. polarization vectors of the emitted boson. Due to presence
of $\sin\phi$ harmonics, the antisymmetrized cross-section directly
probes the imaginary part of the DVMP amplitude as

\begin{equation}
\frac{d^{4}\sigma_{asym}(\phi)}{dt\, d\ln x_{Bj}\, dQ^{2}d\phi}=\frac{d^{4}\sigma(\phi)}{dt\, d\ln x_{Bj}\, dQ^{2}d\phi}-\frac{d^{4}\sigma(-\phi)}{dt\, d\ln x_{Bj}\, dQ^{2}d\phi}\sim\mathcal{S}_{1}^{int}\sin\phi\sim\mathcal{I}m\left(\mathcal{C}^{int}\right).\label{eq:XSec_Asym}
\end{equation}

Another way to access the interference term is based on the isospin
symmetry for the pure DVMP cross-sections~(\ref{eq:isospin_DVMP_1},\ref{eq:isospin_DVMP_2}).
Since the BH correction nearly vanishes on a neutron target, and that
BH cross-sections may be easily calculated, one may directly probe
the interference term and extract the real and imaginary parts of
the DVMP amplitude as, 
\begin{align}
\frac{d^{4}\sigma_{\bar{\nu}\, p\to\mu^{+}\pi^{-}p}^{(int)}}{dt\, d\ln x_{Bj}\, dQ^{2}d\phi} & \approx\frac{d^{4}\sigma_{\bar{\nu}\, p\to\mu^{+}\pi^{-}p}}{dt\, d\ln x_{Bj}\, dQ^{2}d\phi}-\frac{d^{4}\sigma_{\nu\, n\to\mu^{-}\pi^{+}n}}{dt\, d\ln x_{Bj}\, dQ^{2}d\phi}-\frac{d^{4}\sigma_{\bar{\nu}\, p\to\mu^{+}\pi^{-}p}^{(BH)}}{dt\, d\ln x_{Bj}\, dQ^{2}d\phi},\label{eq:XSec_isospin_1}\\
\frac{d^{4}\sigma_{\nu\, p\to\mu^{-}\pi^{+}p}^{(int)}}{dt\, d\ln x_{Bj}\, dQ^{2}d\phi} & \approx\frac{d^{4}\sigma_{\nu\, p\to\mu^{-}\pi^{+}p}}{dt\, d\ln x_{Bj}\, dQ^{2}d\phi}-\frac{d^{4}\sigma_{\bar{\nu}\, n\to\mu^{+}\pi^{-}n}}{dt\, d\ln x_{Bj}\, dQ^{2}d\phi}-\frac{d^{4}\sigma_{\nu\, p\to\mu^{-}\pi^{+}p}^{(BH)}}{dt\, d\ln x_{Bj}\, dQ^{2}d\phi}.\label{eq:XSec_isospin_2}
\end{align}

\section{GPD and DA parametrizations}

\label{sec:Parametrizations}

As was mentioned in the introduction, an essential uncertainty in
the calculations of DVMP originate from the poorly known DAs of the
produced mesons. Only the DAs of pions and $\eta$-mesons have been
tested experimentally, and even in this case the situation remains
rather controversial. The early experiments CELLO and CLEO~\cite{Gronberg:1997fj},
which studied the small-$Q^{2}$ behavior of the form factor $F_{M\gamma\gamma}$,
found it to be consistent with the asymptotic form, $\phi_{as}(z)=6\, z(1-z)$.
Later the BABAR collaboration~\cite{Aubert:2009mc} found a steep
rise with $Q^{2}$ of the form factor $Q^{2}\left|F_{\pi\gamma\gamma}\left(Q^{2}\right)\right|^{2}$
in the large-$Q^{2}$ regime. This observation gave rise to speculations
that the pion DA might have a $z$-dependence quite different from
the asymptotic form~\cite{Polyakov:2009je} (see also the recent
review by Brodsky \emph{et. al.} in~\cite{Brodsky:2011xx,Brodsky:2011yv}).
However, the most recent mesurements by the BELLE collaboration~\cite{Uehara:2012ag}
did not confirm the rapid growth with $Q^{2}$ observed in the BABAR
experiment. As was found in~\cite{Pimikov:2012nm,Bakulev:2012nh}
based on the fits to BELLE, CLEO and CELLO data, the Gegenbauer expansion
coefficients of the pion DA $\phi_{2;\pi}(z)$ are small and give
at most $10\%$ correction for the minus-first moment $\phi_{-1}$
defined in~(\ref{eq:MinusFirstMoment}). Although there are no direct
measurements of the kaon DAs, it is expected that their deviations
from the pion DAs are parametrically suppressed by the quark mass
$m_{s}/GeV$. Numerically this corresponds to a 10-20\% deviation.

For this reason in what follows we assume all the Goldstone DAs to
have the asymptotic form, 
\begin{equation}
\phi_{2;\{\pi,K,\eta\}}(z)\approx\phi_{as}(z)=6\, z(1-z).
\end{equation}
For the decay constants we use the standard values $f_{\pi}\approx93$~MeV,
$f_{K}\approx113$~MeV, and $f_{\eta}\approx f_{K}$.

For GPDs more than a dozen different parametrizations have been proposed
so far~\cite{Diehl:2000xz,Goloskokov:2008ib,Radyushkin:1997ki,Kumericki:2009uq,Kumericki:2011rz,Guidal:2010de,Polyakov:2008aa,Polyakov:2002wz,Freund:2002qf}.
While we neither endorse nor refute any of them, for the sake of concreteness
we select the parametrization~\cite{Goloskokov:2006hr,Goloskokov:2007nt,Goloskokov:2008ib},
which succeeded to describe HERA~\cite{Aaron:2009xp} and JLAB~\cite{Goloskokov:2006hr,Goloskokov:2007nt,Goloskokov:2008ib}
data on electro- and photoproduction of different mesons, so it might
provide a reasonable description of $\nu$DVMP. This parametrization
is based on the Radyushkin's double distribution ansatz. It assumes
additivity of the valence and sea parts of the GPDs, 
\begin{equation}
H(x,\xi,t)=H_{val}(x,\xi,t)+H_{sea}(x,\xi,t),
\end{equation}
which are defined as 
\begin{eqnarray}
H_{val}^{q} & = & \int_{|\alpha|+|\beta|\le1}d\beta d\alpha\delta\left(\beta-x+\alpha\xi\right)\,\frac{3\theta(\beta)\left((1-|\beta|)^{2}-\alpha^{2}\right)}{4(1-|\beta|)^{3}}\, q_{val}(\beta)e^{\left(b_{i}-\alpha_{i}\ln|\beta|\right)t};\\
H_{sea}^{q} & = & \int_{|\alpha|+|\beta|\le1}d\beta d\alpha\delta\left(\beta-x+\alpha\xi\right)\,\frac{3\, sgn(\beta)\left((1-|\beta|)^{2}-\alpha^{2}\right)^{2}}{8(1-|\beta|)^{5}}\, q_{sea}(\beta)e^{\left(b_{i}-\alpha_{i}\ln|\beta|\right)t};
\end{eqnarray}
and $q_{val}$ and $q_{sea}$ are the ordinary valence and sea components
of PDFs. The coefficients $b_{i}$, $\alpha_{i}$, as well as the
parametrization of the input PDFs $q(x),\,\Delta q(x)$ and pseudo-PDFs
$e(x),\,\tilde{e}(x)$ (corresponding to the forward limit of the
GPDs $E,\,\tilde{E}$), are discussed in~\cite{Goloskokov:2006hr,Goloskokov:2007nt,Goloskokov:2008ib}.
The unpolarized PDFs $q(x)$ within the range of $Q^{2}\lesssim40$~GeV$^{2}$
roughly coincide with the CTEQ PDFs. Notice that in this model the
sea is flavor symmetric for asymptotically large $Q^{2}$, 
\begin{equation}
H_{sea}^{u}=H_{sea}^{d}=\kappa\left(Q^{2}\right)H_{sea}^{s},\label{eq:SeaFlavourSymmetry}
\end{equation}
where 
\begin{equation}
\kappa\left(Q^{2}\right)=1+\frac{0.68}{1+0.52\ln\left(Q^{2}/Q_{0}^{2}\right)},\quad Q_{0}^{2}=4\, GeV^{2}.
\end{equation}

The equality of the sea components for $u$ and $d$ quarks in~(\ref{eq:SeaFlavourSymmetry})
should be considered as a rough approximation, since in the forward
limit $\bar{d}\not=\bar{u}$ was firmly established by the E866/NuSea
experiment~\cite{Hawker:1998ty}. For this reason the predictions
made with this parametrization of GPDs for the $p\rightleftarrows n$
transitions in the region $x_{Bj}\in(0.1...0.3)$ might slightly underestimate
the data.

The Dirac and Pauli form factors $F_{1}(t),\, F_{2}(t)$ are extracted
from GPDs in the standard way, 
\begin{align}
F_{1}^{em}(t) & =\sum_{q}e_{q}\int_{-1}^{1}dx\, H_{q}(x,\xi,t),\\
F_{2}^{em}(t) & =\sum_{q}e_{q}\int_{-1}^{1}dx\, E_{q}(x,\xi,t).
\end{align}

\section{Numerical results and discussion}

\label{sec:Results}

In this section we perform numerical analysis of the electromagnetic
corrections to the processes listed in Table~\ref{tab:DVMP_amps},
relying on the GPDs described in the previous section. At small $Q^{2}$
the angular harmonics are small and the cross-section is dominated
by the angular-independent DVMP contribution. Therefore, it is convenient
to normalize all the coefficients to DVMP cross-section, 
\begin{equation}
\frac{d^{4}\sigma}{dt\, d\ln x_{B}\, dQ^{2}d\phi}=\frac{d^{4}\sigma^{(DVMP)}}{dt\, d\ln x_{B}\, dQ^{2}d\phi}\left(\sum_{n=0}^{2}c_{n}\cos n\phi+s_{1}\sin\phi\right).\label{xsect}
\end{equation}
Notice that in the limit $\alpha_{em}\to0$, no BH corrections are
possible, the coefficient $c_{0}=1$, and all other coefficients vanish.

\begin{figure}[htb]
\includegraphics[scale=0.4]{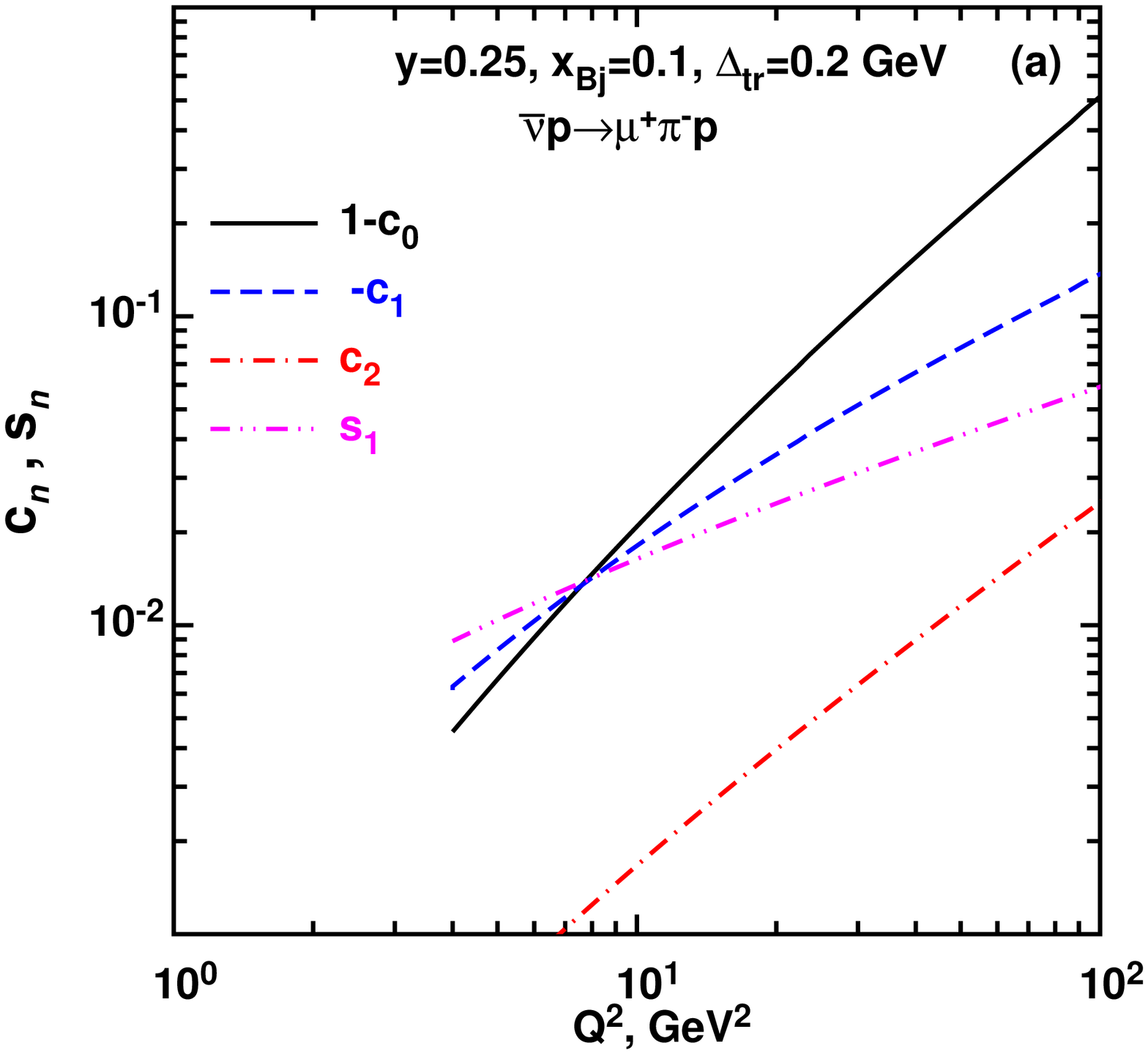}\qquad{}\includegraphics[scale=0.4]{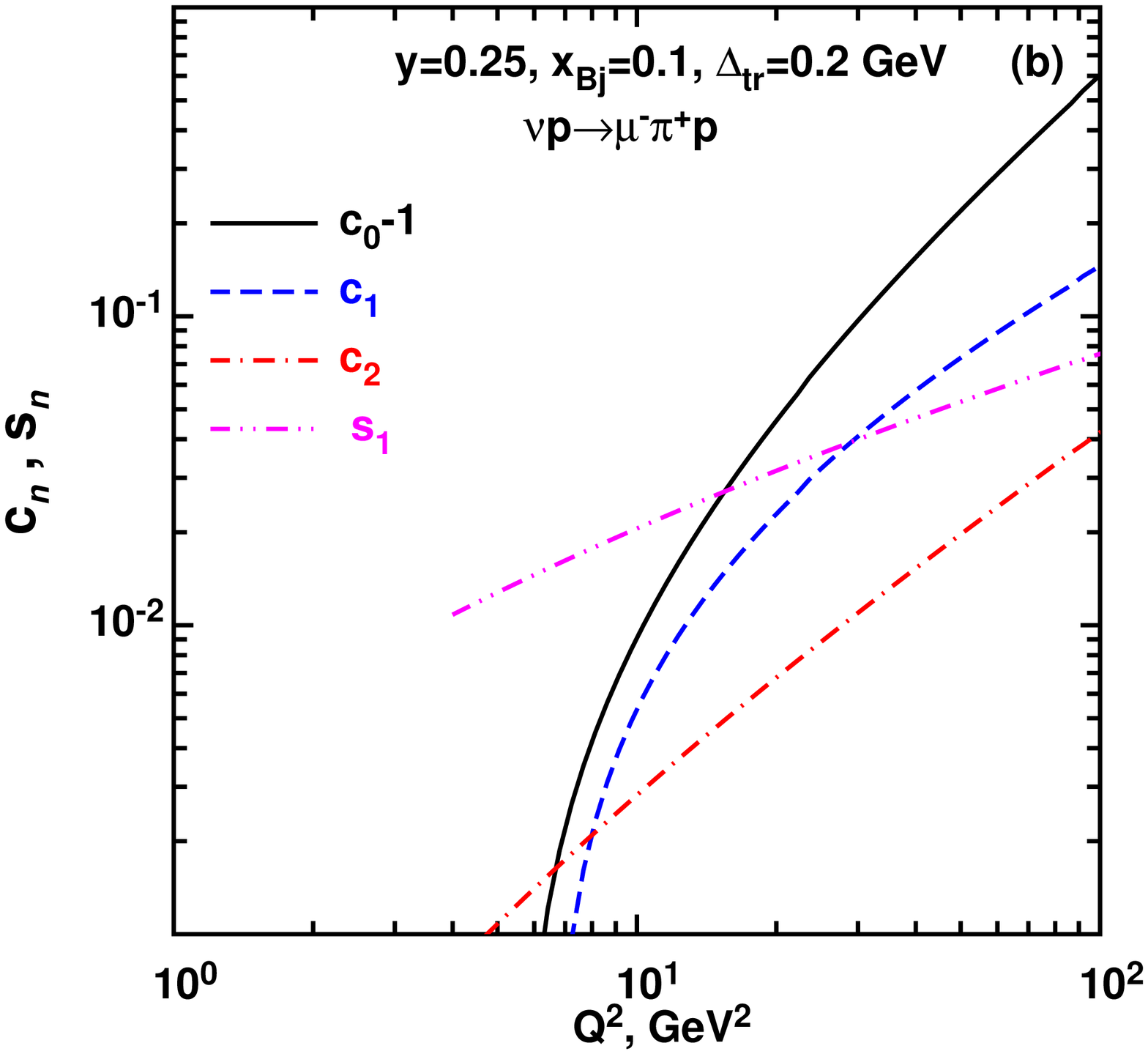}\\
 \includegraphics[scale=0.4]{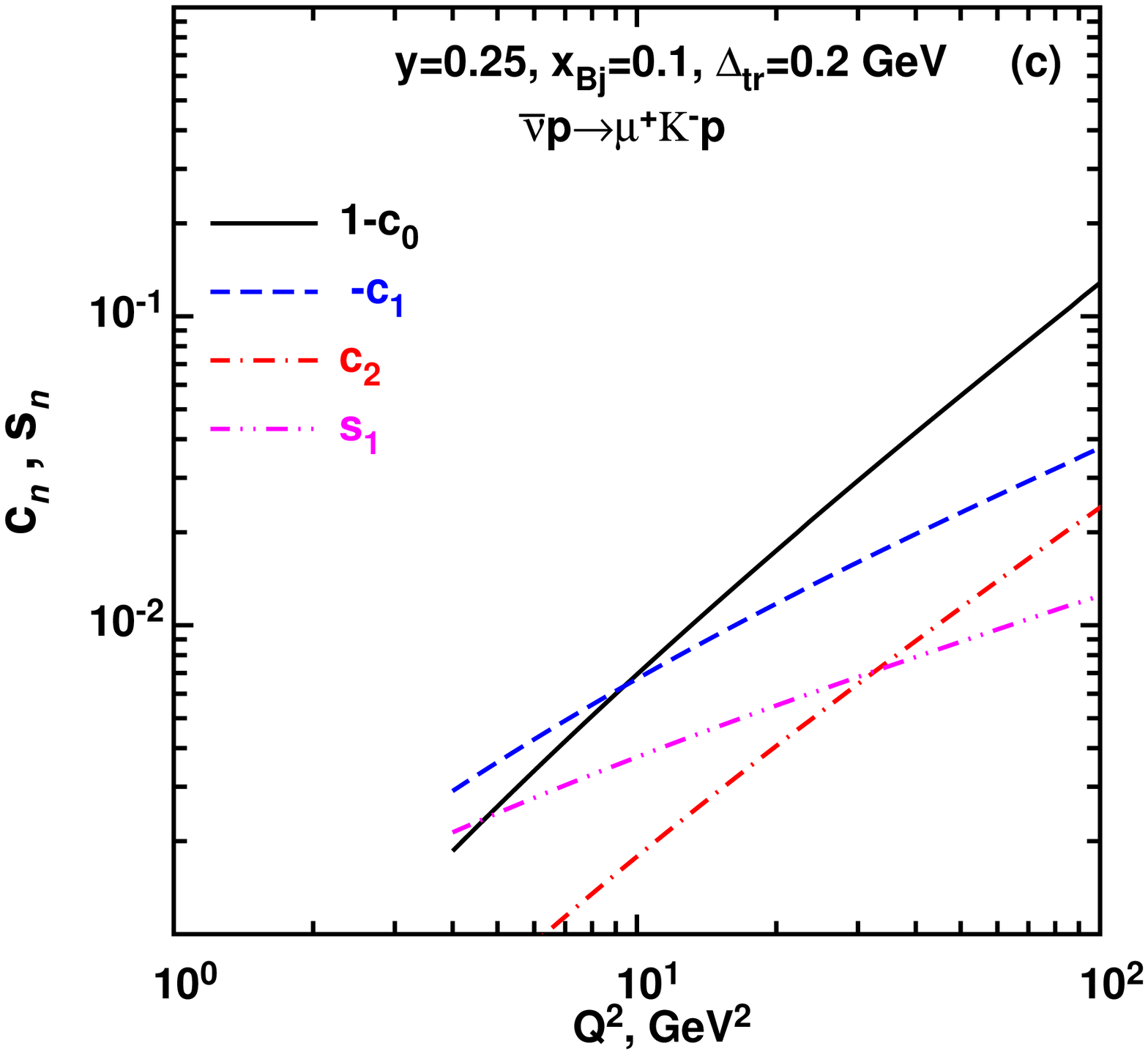}\qquad{}\includegraphics[scale=0.4]{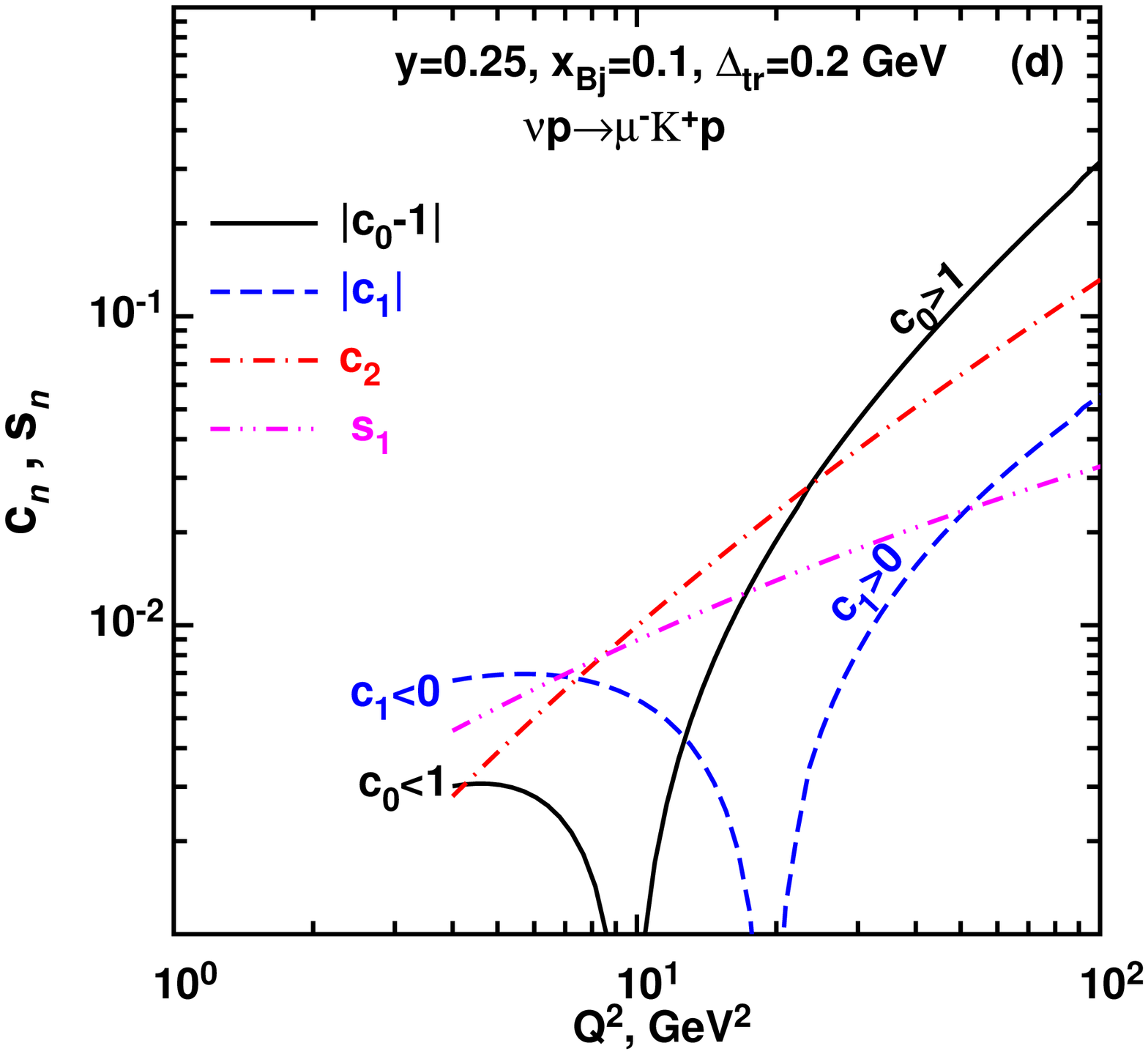}
\caption{\label{fig:DVMP-Q2}(color online) $Q^{2}$-dependence of the BH correction
to the $\nu$DVMP process. }
\end{figure}

The results for the $Q^{2}$-dependence of the relative BH corrections
to the neutrino-DVMP cross section for pions and kaons are presented
in Figure~\ref{fig:DVMP-Q2}. 

We see that the isotropic part of the BH correction $1-c_{0}$ steeply
rises in all channels from few percent or less at $Q^{2}\lesssim10\GeV^{2}$
up to few tens of percent at $Q^{2}\sim100\GeV^{2}$. 
It behaves like $1-c_{0}\propto Q^{2}$ modulo logarithmic corrections.
As a result, the cross section is reduced about twice compared to
the DVMP contribution. The asymmetry $s_{1}$ also rises with $Q^{2}$
and reaches about $15\,\%$ at $Q^{2}=100$~GeV$^{2}$. Notice that
some of the coefficients (e.g. $c_{0}-1$, $c_{1}$) have nodes for
$\pi^{+}/K^{+}$ production, while they are absent for $\pi^{-}/K^{-}$.
The reason is purely algebraic: as one can see from the Table~\ref{tab:DVMP_amps},
for $\pi^{+}/K^{+}$ the large $s$-channel coefficient function $c_{-}$
is convoluted with the small $d/s$ -quark GPD, whereas the small
$u$-channel coefficient function $c_{+}$ is convoluted with the
large $u$-quark GPD. This produces a node in the real part of the
DVMP amplitude, because the real parts of the two contributions have
opposite signs. Such a node is absent for negatively charged mesons,
because the ``large'' $u$-quark GPD is convoluted with the ``large''
$c_{-}$. The full DVMP cross section has no nodes, because it gets
a large contribution from the imaginary part, which homogeneously
depends on $Q^{2}$ (the coefficient $s_{1}$, which probes the imaginary
part, has no nodes). The difference between the Cabibbo suppressed
and allowed processes comes from the sensitivity to different flavor
combinations of GPDs in the corresponding DVMP amplitude.

The terms $c_{0}-1$ and $c_{1}$ in Eqn.~(\ref{xsect}) are dominated
by the interference of the DVMP and BH amplitudes, therefore they
have different signs for $\pi^{+}$ and $\pi^{-}$ (and $K^{+}$ and
$K^{-}$). The term $c_{2}$ gets contribution only from BH process,
so it always has the same sign. The $s_{1}$-term does not change
its sign under the $C$-conjugation in the lepton part, because it
originates from the $P$-odd interference between the vector and axial
vector currents.

\begin{figure}[htb]
\includegraphics[scale=0.4]{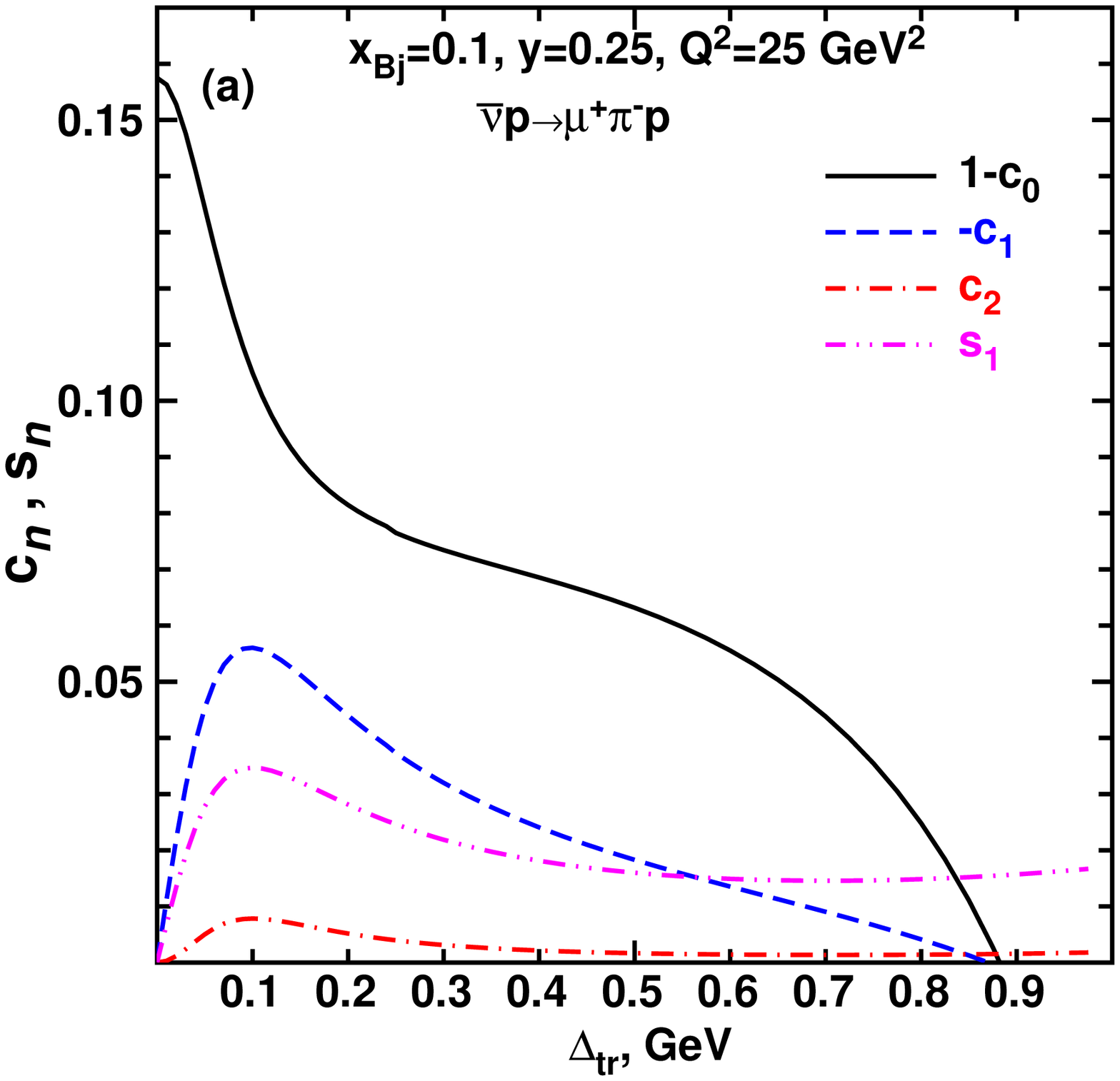}\qquad{}\includegraphics[scale=0.4]{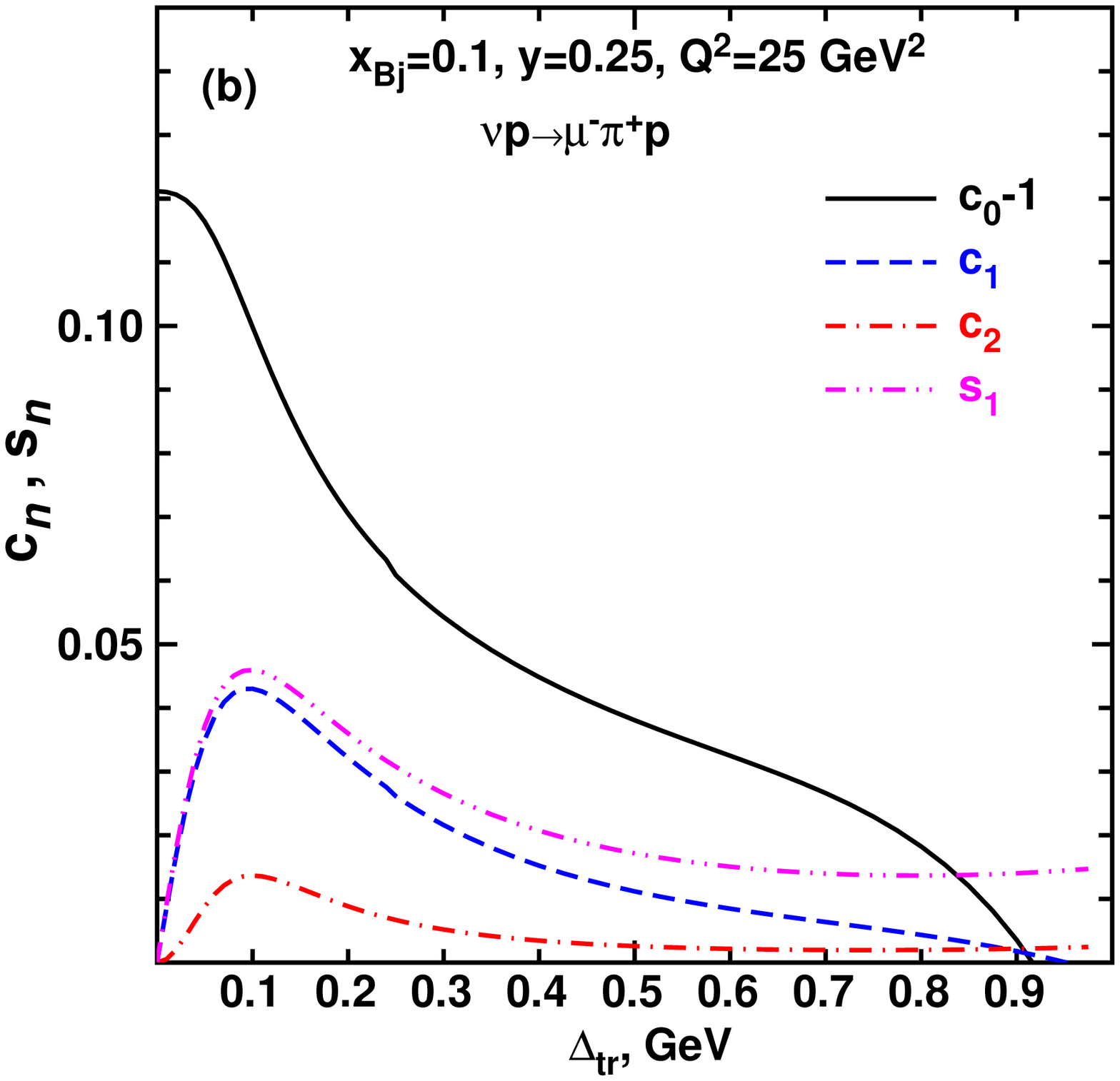}\\
 \includegraphics[scale=0.4]{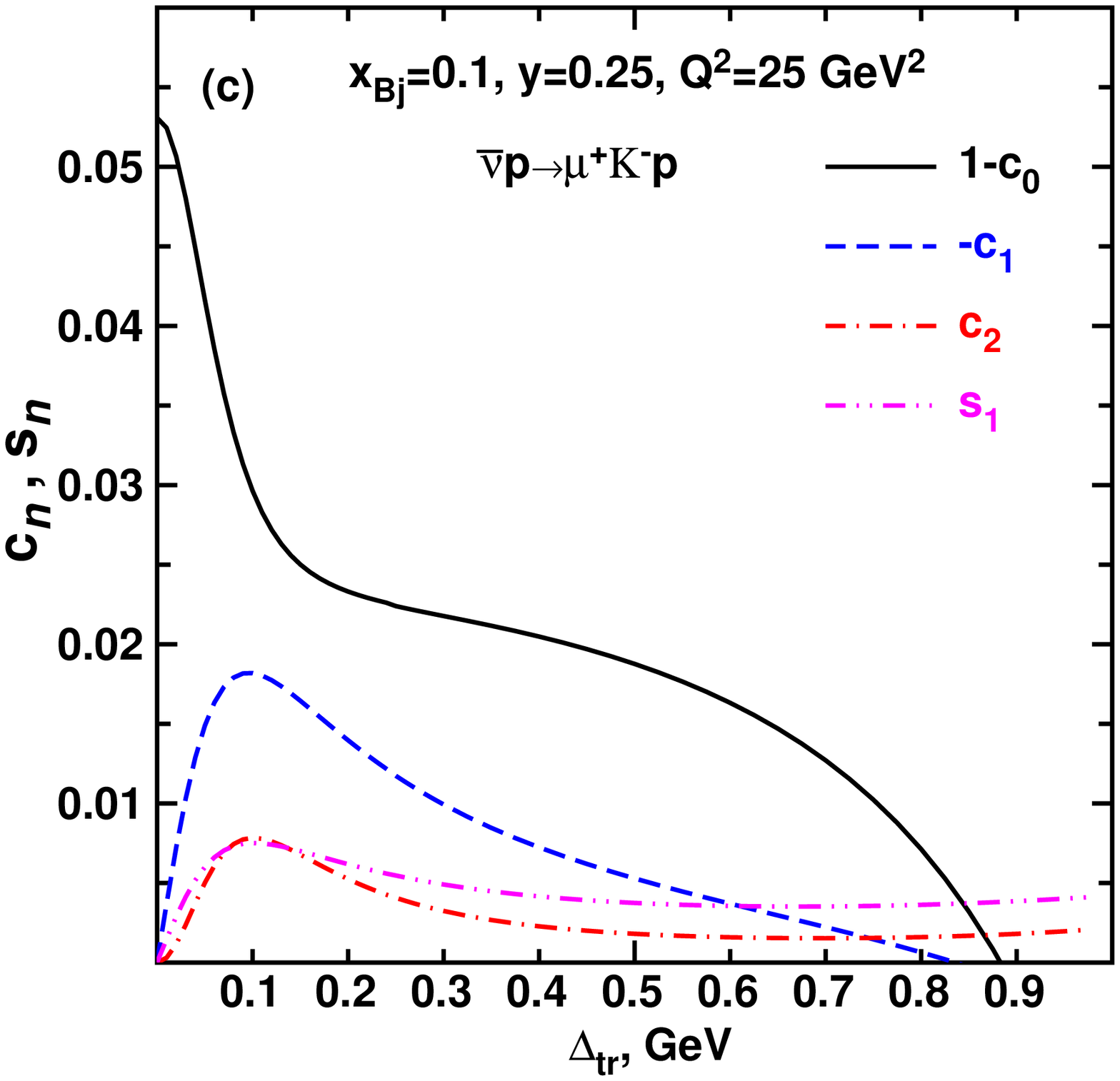}\qquad{}\includegraphics[scale=0.4]{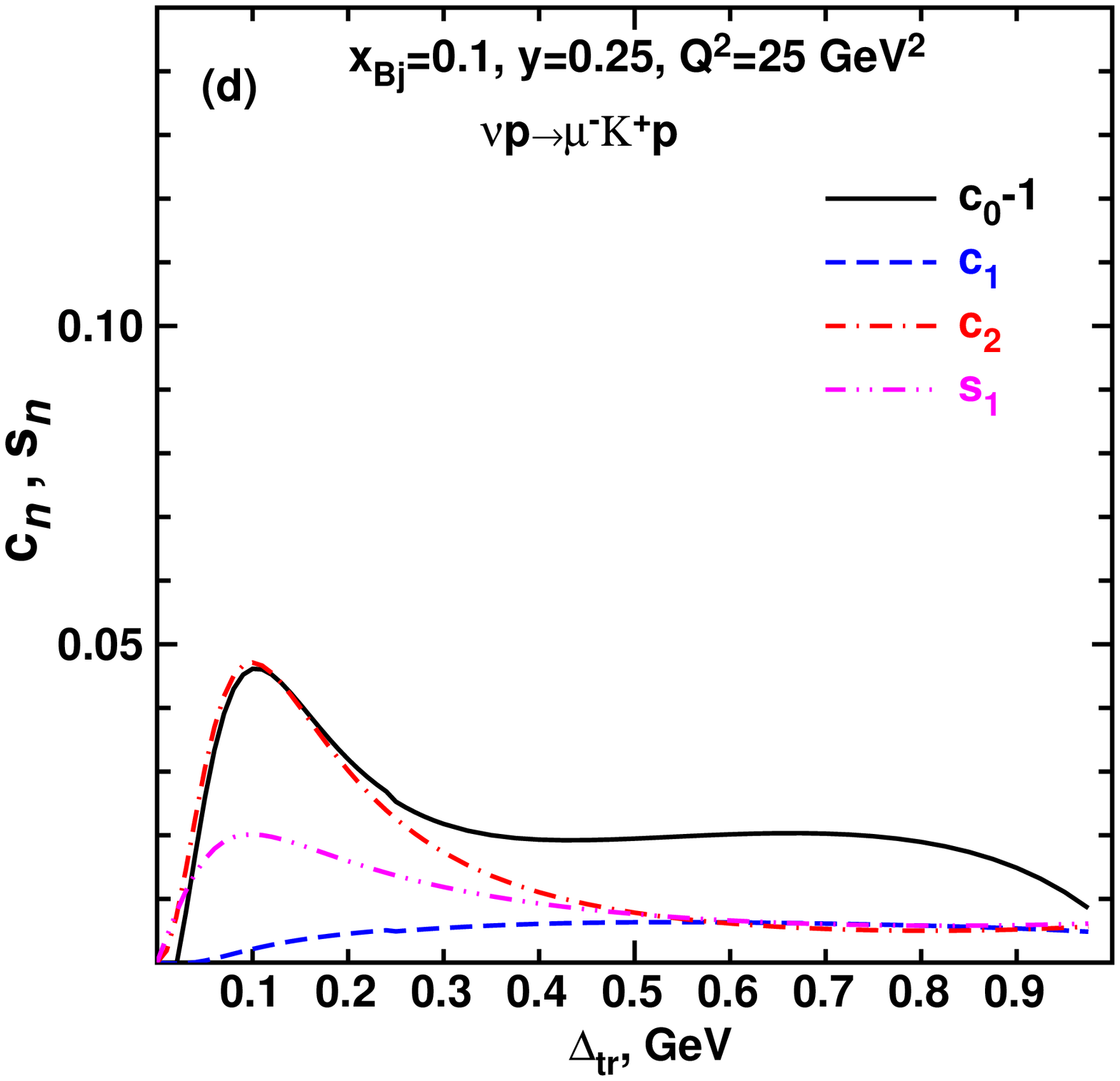}
\caption{\label{fig:DVMP-pT}(color online) $\Delta_{\perp}$-dependence of
the BH correction to the $\nu$DVMP process. }
\end{figure}

The results for $\Delta_{\perp}$-dependence of the BH corrections
are depicted in Figure~\ref{fig:DVMP-pT}. 

It shows that there is a qualitative difference between $c_{0}$ and
other angular harmonics. The coefficient $c_{0}$ reaches its maximum
at $\Delta_{\perp}=0$ due to the $1/t$ behavior of the BH cross
section. In contrast, the angular harmonics $c_{1},\, c_{2},\, s_{1}$
vanish at small $\Delta_{\perp}$ due the $K$-factors in front of
them. As a consequence, the harmonics reach their maxima at $\Delta_{\perp}\sim0.1$~GeV.

\begin{figure}
\includegraphics[scale=0.4]{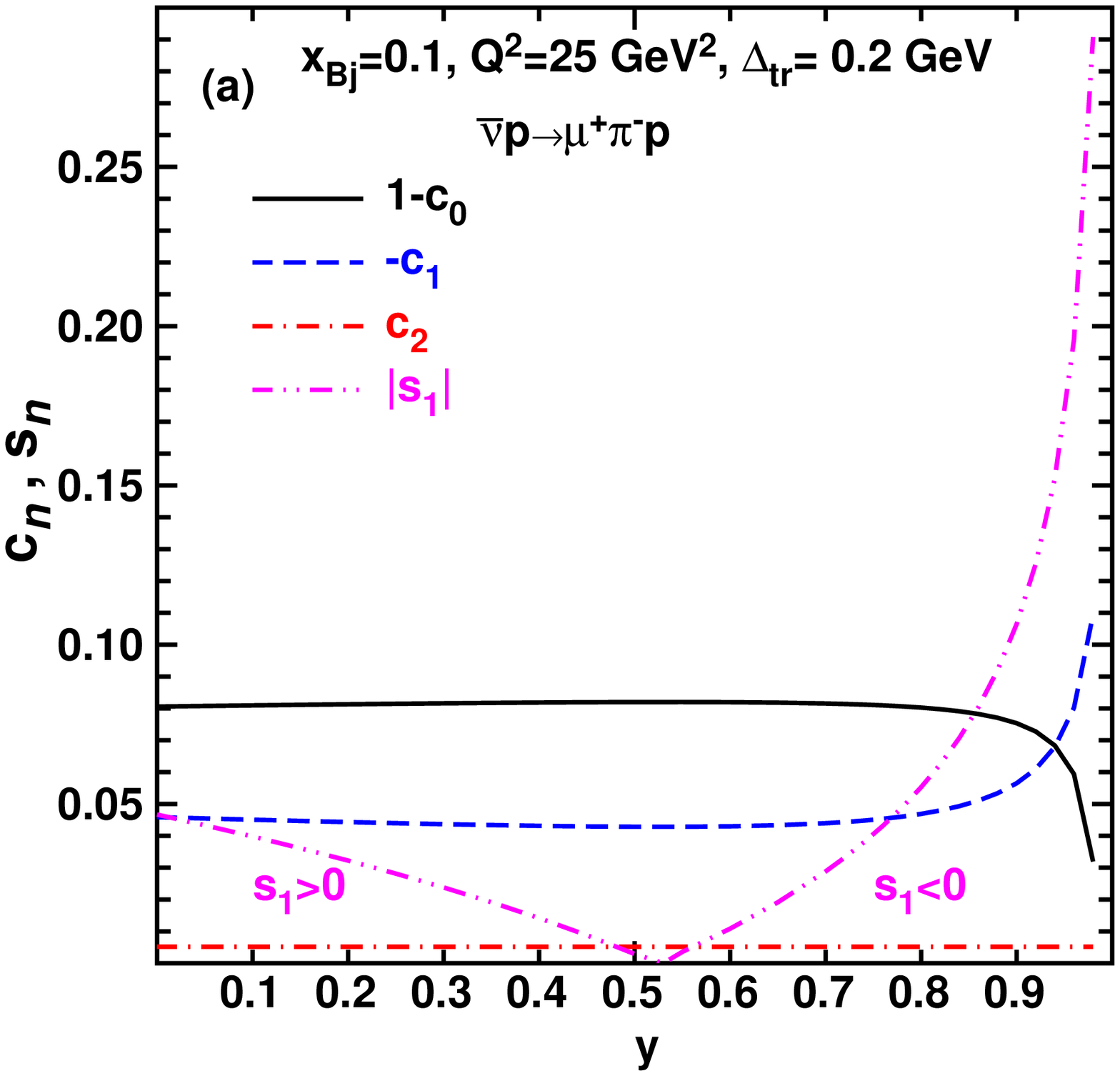}\qquad{}\includegraphics[scale=0.4]{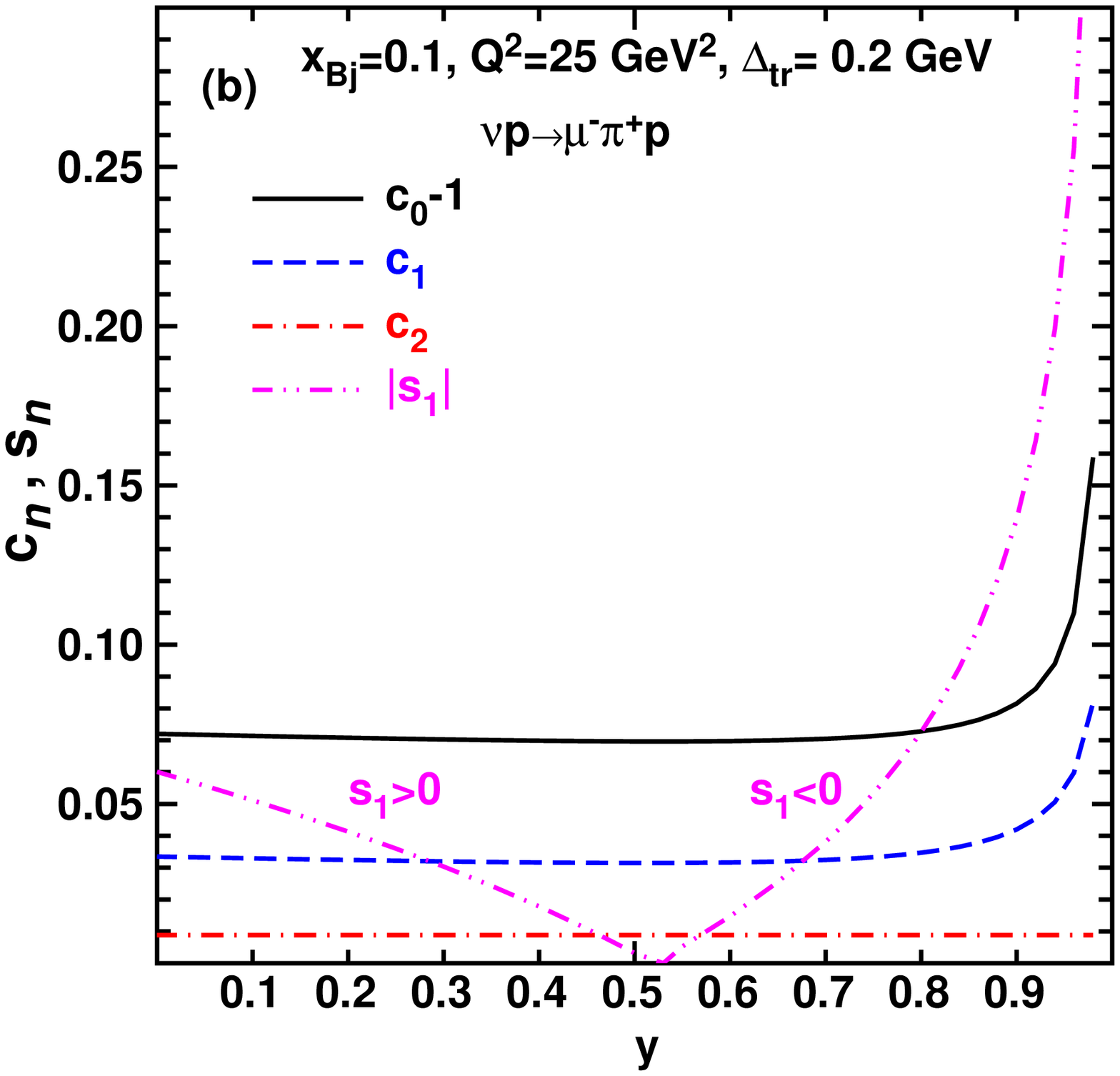}\\
 \includegraphics[scale=0.4]{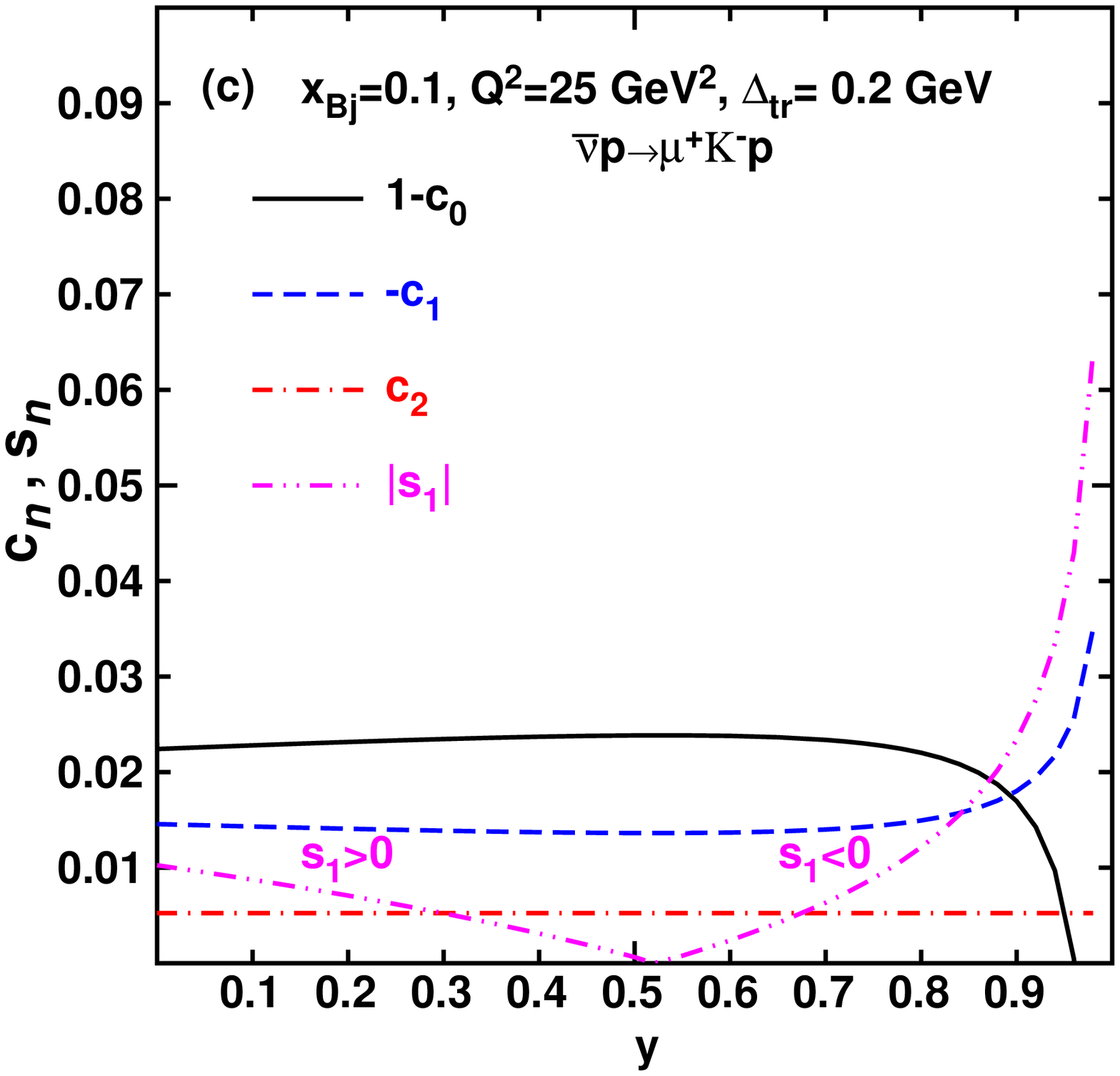}\qquad{}\includegraphics[scale=0.4]{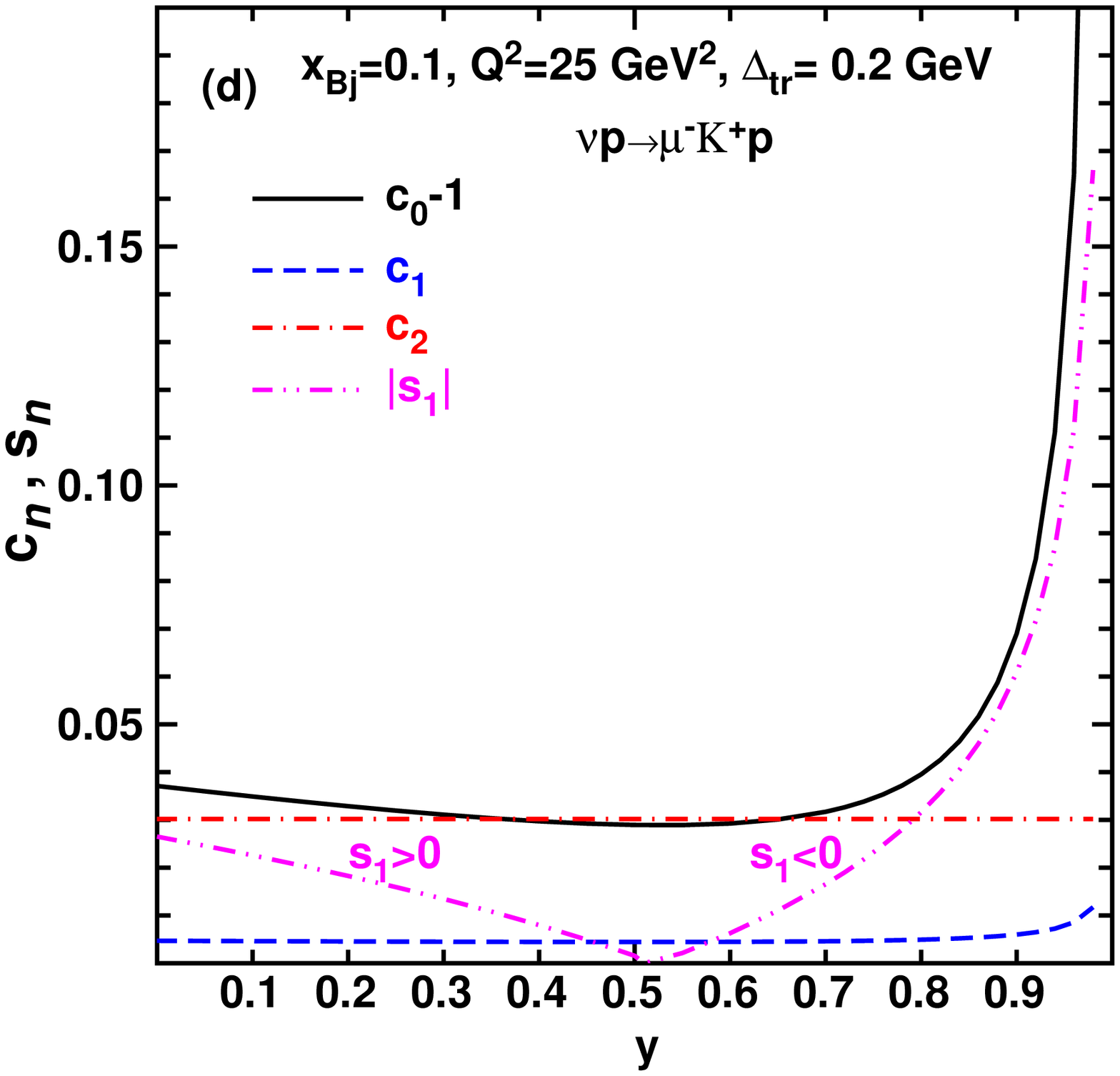}
\caption{\label{fig:DVMP-y}(color online) $y$-dependence of the BH correction
to the $\nu$DVMP process. }
\end{figure}

The results for the angular harmonic coefficients vs the elasticity
parameter $y=P\cdot q/P\cdot k=\nu/E_{\nu}$ are presented in Figure~\ref{fig:DVMP-y}.

The coefficient $c_{2}$ does not depend on $y$ at all due to exact
cancellation of the pre-factors $1-y-y^{2}\epsilon^{2}/4$ in $K^{2}$
and the pre-factor in the DVMP cross-section Eqn.~(\ref{eq:XSec_DVMP}).
The harmonics $c_{0}$ and $c_{1}$ have a mild dependence on $y$,
except in the region $y\sim1$, where they blow up, because the DVMP
cross-section~(\ref{eq:XSec_DVMP}) is suppressed there by the factor
$\sim1-y-y^{2}\epsilon^{2}/4$, whereas the harmonics in~(\ref{eq:XSec_BH},\ref{eq:XSec_Int})
are suppressed at most as $K\propto\sqrt{1-y-y^{2}\epsilon^{2}/4}.$
The harmonic $s_{1}$ in accordance with Eqn.~(\ref{eq:Int_S1})
in the kinematics $x_{B}\ll1,\,|t|\ll m_{N}^{2}$ is proportional
to $\sim(2y-1)F_{1}(t)\mathcal{I}m\,\mathcal{H}$ and has a node near
$y\approx0.5$.

\section{Summary}

In this paper we studied the electromagnetic Bethe-Heitler corrections
to neutrino-induced deeply virtual meson production. We found these
corrections to fall with $Q^{2}$ less steeply compared with the $\nu$DVMP
cross section, so they tend to become a dominant mechanism in the
Bjorken limit of $Q^{2}\to\infty$. Besides, they are enhanced at
small-$t$ due to the $t$-channel Coulomb pole $\propto1/t$. Remarkably,
these corrections generate an angular correlation between the lepton
and hadron scattering planes. Similar to the BH corrections in DVCS,
some angular harmonics are sensitive to the real or imaginary parts
of the DVMP amplitude (see (\ref{eq:Int_C0}-\ref{eq:Int_S1})). Notice
that the appearance of such angular dependence was also predicted
in~\cite{Goldstein:2009in}, however there it appears due to interference
of the longitudinally and transversely polarized charged bosons, which
is a twist-three effect. In contrast, our result is a twist-two effect,
and as one can see from Figure~\ref{fig:DVMP-Q2}, it is not suppressed
for asymptotically large $Q^{2}$.

Numerically, the BH corrections are subject to the interplay between
the suppression factor $\sim\alpha_{em}$ and the relative enhancement,
which is as large as $\sim Q^{2}/t$ for some harmonics. In the kinematics
of the \textsc{Minerva} experiment, the BH contribution for the proton
target represents a few percent correction and thus is important for
precision tests of the GPD parametrizations. At $Q^{2}\sim100$ GeV$^{2}$,
which can be accessed in future neutrino experiments, these corrections
are expected to become on par with the DVMP contribution. For a neutron
target, these corrections are two orders of magnitude smaller than
for a proton and can be neglected up to very high $Q^{2}$. Combining
this fact with isospin symmetry of the DVMP amplitude, we construct
combinations of the cross-sections Eqs.~(\ref{eq:XSec_isospin_1},\ref{eq:XSec_isospin_2}),
which are sensitive only to the interference term.

The electromagnetic corrections discussed in this paper are important
only in neutrino-induced DVMP: in the case of electron-induced processes
$ep\to ep\, M$ the BH corrections are suppressed by the factor $\left(G_{F}^{2}Q^{4}\right)$,
and are negligibly small. We provide a computational code, which can
be used for evaluation of the cross-sections relying on different
GPD models.

\appendix

\section{Evaluation of the $\nu$DVMP and BH cross-sections}

\label{sec:DVMP_Xsec-App} In this section we present some technical
details of the evaluation of diagrams (a-c) in the Figure~\ref{fig:DVMPLT}.
The calculation of the diagram (a) in Figure~\ref{fig:DVMPLT} is
rather straightforward and yields for the amplitude of the process~\cite{Mankiewicz:1998kg,Vanderhaeghen:1998uc,Kopeliovich:2012dr}
\begin{eqnarray}
T_{a} & =\frac{8\pi i}{9}\frac{G_{F}}{\sqrt{2}}\frac{\bar{\mu}\left(k_{\mu}\right)\hat{\epsilon}_{L}^{*}(q)\left(1-\gamma_{5}\right)\nu\left(k_{\nu}\right)}{\left(1+Q^{2}/M_{W}^{2}\right)} & \frac{\alpha_{s}}{Q}\phi_{-1}\sum_{\Gamma}\mathcal{H}_{M}^{\Gamma}\bar{N}\left(p_{2}\right)\Gamma N\left(p_{1}\right),\label{eq:T_def-1}
\end{eqnarray}
where $\bar{\mu}(k)$ and $\nu(k)$ are the spinors of the final muon
and initial neutrino; $\epsilon(k)$ is the polarization vector of
the photon; $N(p),\,\bar{N}(p)$ are the spinors of the initial/final
state baryons; $\phi_{M}(z)$ is the DA of the produced meson; $f_{M}$
is the decay constant of the meson $M$; subscript index for each
momentum in Eqn.~(\ref{eq:T_def-1}) and in what follows shows to
which particle it corresponds; $\sum_{\Gamma}H_{M}^{\Gamma}\bar{N}\left(p_{2}\right)\Gamma N\left(p_{1}\right)$
is a symbolic notation for summation of all the leading twist GPDs
contributions (defined below); and $\mathcal{H}_{M}^{\Gamma}$ are
the convolutions of the GPDs $H_{\Gamma}$ of the target with the
proper coefficient function. Currently, the amplitude of the DVMP
is known up to the NLO accuracy~\cite{Ivanov:2004zv,Diehl:2007hd}.
Extension of the analysis of~\cite{Vanderhaeghen:1998uc,Mankiewicz:1998kg}
to neutrinos is straightforward. In contrast to electro-production,
due to the $V-A$ structure, the amplitudes acquire contributions
from both the unpolarized and helicity flip GPDs.

Four GPDs, $H,\, E,\,\tilde{H}$ and $\tilde{E}$ contribute to this
process in the leading twist. They are defined as 
\begin{eqnarray}
\frac{\bar{P}^{+}}{2\pi}\int dz\, e^{ix\bar{P}^{+}z}\left\langle A\left(p_{2}\right)\left|\bar{\psi}_{q'}\left(-\frac{z}{2}\right)\gamma_{+}\psi_{q}\left(\frac{z}{2}\right)\right|A\left(p_{1}\right)\right\rangle  & = & \left(H_{q}\left(x,\xi,t\right)\bar{N}\left(p_{2}\right)\gamma_{+}N\left(p_{1}\right)\right.\label{eq:H_def-1}\\
 &  & \left.+\frac{\Delta_{k}}{2m_{N}}E_{q}\left(x,\xi,t\right)\bar{N}\left(p_{2}\right)i\sigma_{+k}N\left(p_{1}\right)\right)\nonumber \\
\frac{\bar{P}^{+}}{2\pi}\int dz\, e^{ix\bar{P}^{+}z}\left\langle A\left(p_{2}\right)\left|\bar{\psi}_{q'}\left(-\frac{z}{2}\right)\gamma_{+}\gamma_{5}\psi_{q}\left(\frac{z}{2}\right)\right|A\left(p_{1}\right)\right\rangle  & = & \left(\tilde{H}_{q}\left(x,\xi,t\right)\bar{N}\left(p_{2}\right)\gamma_{+}\gamma_{5}N\left(p_{1}\right)\right.\label{eq:Htilde_def-1}\\
 &  & \left.+\frac{\Delta_{+}}{2m_{N}}\tilde{E}_{q}\left(x,\xi,t\right)\bar{N}\left(p_{2}\right)N\left(p_{1}\right)\right),\nonumber 
\end{eqnarray}
where $\bar{P}=p_{1}+p_{2}$, $\Delta=p_{2}-p_{1}$ and $\xi=-\Delta^{+}/2\bar{P}^{+}\approx x_{Bj}/(2-x_{Bj})$
(see e.g.~\cite{Goeke:2001tz} for the details of kinematics). In
what follows we assume that the target $A$ is either a proton or
a neutron. Since in neutrino experiments the target cannot be polarized
due to its large size, it makes no sense to discuss the transversity
GPDs $H_{T},\, E_{T},\,\tilde{H}_{T},\,\tilde{E}_{T}$. We also ignore
the contributions of gluons in this paper because in the current neutrino
experiments the region of small $x_{Bj}\ll1$ but very high $Q^{2}$,
is hardly accessible, so the amplitude~(\ref{eq:T_def-1}) simplifies
to 
\begin{eqnarray}
T_{a} & = & \frac{8\pi i}{9}\frac{G_{F}}{\sqrt{2}}\frac{\bar{\mu}\left(k_{\mu}\right)\hat{\epsilon}_{L}(q)\nu\left(k_{\nu}\right)}{\left(1+Q^{2}/M_{W}^{2}\right)}\frac{\alpha_{s}}{Q}\left(\int dz\frac{\phi_{M}(z)}{z}\right)\left[\left(\tilde{\mathcal{H}}_{M}\bar{N}\left(p_{2}\right)\gamma_{+}\gamma_{5}N\left(p_{1}\right)+\frac{\Delta_{+}}{2m_{N}}\tilde{\mathcal{E}}_{M}\bar{N}\left(p_{2}\right)\gamma_{5}N\left(p_{1}\right)\right)\right.\label{eq:T_LT-1}\\
 & + & \left.\left(\mathcal{H}_{M}\bar{N}\left(p_{2}\right)\gamma_{+}N\left(p_{1}\right)+\frac{\Delta_{k}}{2m_{N}}\mathcal{E}_{M}\bar{N}\left(p_{2}\right)i\sigma_{+k}N\left(p_{1}\right)\right)\right],\nonumber 
\end{eqnarray}

In Table~\ref{tab:DVMP_amps} the corresponding amplitudes are listed
for each final state $M$. The DVMP part of the corresponding neutrino
cross-section for charged currents is given by~(\ref{eq:XSec_DVMP}).

In the leading order in $Q^{2}$ both BH diagrams Figure~\ref{fig:DVMPLT}~(b,c)
are dominated by longitudinally polarized photons. Nevertheless, as
was mentioned in Section~\ref{sec:DVMP_Xsec}, we evaluate the BH
contribution exactly, because various angular harmonics, suppressed
by $\sim\Delta_{\perp}/Q$, get contribution from the transverse components,
which is of the same order as the longitudinal result. Only after
that we make expansion in $1/Q^{2}$.

The dipole scattering amplitude, which contributes to the diagram
Figure~\ref{fig:DVMPLT}~(b), has the form
\begin{equation}
\mathcal{A}_{\mu\nu}^{ab}\left(q,\Delta\right)=\frac{1}{f_{\pi}}\int d^{4}x\, e^{-iq\cdot x}\left\langle 0\left|\left(V_{\mu}^{a}(x)-A_{\mu}^{a}(x)\right)J_{\nu}^{em}(0)\right|\pi^{b}\left(q-\Delta\right)\right\rangle ,\label{A5}
\end{equation}
where $V_{\mu}^{a}(x)$ and $A_{\mu}^{a}(x)$ are the vector and axial-vector
isovector currents. Notice that the amplitude $\mathcal{A}_{\mu\nu}$
should not be interpreted as a pion form factor, because: (i) the
virtuality is large; (ii) the insertion of the pion state between
$A_{\mu}^{5}$ and $J_{\nu}^{em}$ leads to $\mathcal{A}_{\mu\nu}^{ab}\sim q_{\mu}$,
which vanishes when is multiplied by an on-shell lepton current.

We evaluated~(\ref{A5}) in pQCD in the collinear approximation.
This is justified in Bjorken kinematics by the high virtuality of
the charged boson, so we assume that the dominant contribution comes
from the leading twist-2 pion DA. The result reads, 
\begin{align}
\mathcal{A}_{\mu\nu}^{\pi^{+}}\left(q,\Delta\right) & =\frac{1}{4}\left(g_{\mu\nu}f_{0}+\left(q_{\mu}n_{\nu}+q_{\nu}n_{\mu}\right)f_{1}+\left(\Delta_{\mu}n_{\nu}+\Delta_{\nu}n_{\mu}\right)f_{2}-i\epsilon_{\mu\nu\beta\gamma}n_{\beta}q_{\gamma}g_{1}-i\epsilon_{\mu\nu\beta\gamma}n_{\beta}\Delta_{\gamma}g_{2}\right);\\
f_{0} & =\int dz\,\phi_{M}(z)\frac{\left(-2z\bar{z}+\left(1-2z\bar{z}\right)t/Q^{2}\right)n\cdot q-\left(\left(1-2z\bar{z}\right)-2z\bar{z}t/Q^{2}\right)n\cdot\Delta}{x_{B}\left(z-\bar{z}t/Q^{2}\right)\left(\bar{z}-zt/Q^{2}\right)}\approx2\phi_{-1}+\mathcal{O}\left(\frac{m_{N}^{2}}{Q^{2}},\frac{t}{Q^{2}}\right);\\
f_{1} & =\int dz\,\phi_{M}(z)\frac{2z\bar{z}-\left(1-2z\bar{z}\right)t/Q^{2}}{x_{B}\left(z-\bar{z}t/Q^{2}\right)\left(\bar{z}-zt/Q^{2}\right)}\approx\frac{2}{x_{B}}+\mathcal{O}\left(\frac{m_{N}^{2}}{Q^{2}},\frac{t}{Q^{2}}\right);\\
f_{2} & =\int dz\,\phi_{M}(z)\frac{\left(1-2z\bar{z}\right)-2z\bar{z}t/Q^{2}}{x_{B}\left(z-\bar{z}t/Q^{2}\right)\left(\bar{z}-zt/Q^{2}\right)}\approx\frac{2}{x_{B}}\left(\phi_{-1}-1\right)+\mathcal{O}\left(\frac{m_{N}^{2}}{Q^{2}},\frac{t}{Q^{2}}\right);\\
g_{1} & =\frac{1}{3\, q\cdot n}\int dz\,\phi_{M}(z)\frac{-2z\bar{z}+\left(1-2z\bar{z}\right)t/Q^{2}}{x_{B}\left(z-\bar{z}t/Q^{2}\right)\left(\bar{z}-zt/Q^{2}\right)}\approx-\frac{2}{3x_{B}}+\mathcal{O}\left(\frac{m_{N}^{2}}{Q^{2}},\frac{t}{Q^{2}}\right)\\
g_{2} & =-\frac{1}{3\, q\cdot n}\int dz\,\phi_{M}(z)\frac{1-2\, z\bar{z}-2\, z\bar{z}\, t/Q^{2}}{x_{B}\left(z-\bar{z}t/Q^{2}\right)\left(\bar{z}-zt/Q^{2}\right)}\approx\frac{2}{3x_{B}}\left(1-\phi_{-1}\right)+\mathcal{O}\left(\frac{m_{N}^{2}}{Q^{2}},\frac{t}{Q^{2}}\right)
\end{align}
Here $p_{\mu}$ and $n_{\mu}$ are the positive and the negative direction
light-cone vectors respectively. The plus-components of $q$ and $\Delta$
have the form, 
\begin{align}
n\cdot q & =\frac{Q^{2}\left(1-\sqrt{1+\epsilon^{2}}\right)}{2m_{N}^{2}x_{B}}\approx-x_{B}+\mathcal{O}\left(\epsilon^{2}\right);\\
n\cdot\Delta & =\frac{\left(x_{B}-t/m_{N}^{2}\right)\left(1-\sqrt{1+\epsilon^{2}}\right)-x_{B}\left(2x_{B}+1-\sqrt{1+\epsilon^{2}}\right)t/Q^{2}}{\left(1-\sqrt{1+\epsilon^{2}}+\epsilon^{2}\right)};\\
 & \approx-x_{B}+\mathcal{O}\left(\epsilon^{2},\,\frac{t}{Q^{2}}\right),\nonumber 
\end{align}
where $\epsilon=2m_{N}x_{B}/Q$. In what follows we encounter the
combination $4-f_{0}-x_{B}(f_{1}-f_{2})$, for which we need to make
expansion up to $\mathcal{O}(Q^{-2})$. While separately the series
expansion coefficients for each factor $f_{i}$ have non-integrable
singularities $\sim z^{-2}\bar{z}^{-2}$, which signal a sensitivity
to the transverse degrees of freedom and presence of the non-analytic
terms $\sim\ln(Q^{2}/|t|)/Q^{2}$, in the above-mentioned combination,
these terms cancel each other resulting in 
\begin{equation}
f_{0}+x_{B}\left(f_{1}-f_{2}\right)\approx4+2\phi_{-1}\frac{t\left(1+x_{B}\right)-2m_{N}^{2}x_{B}^{2}}{Q^{2}}-\frac{2t\left(1+x_{B}\right)-6m_{N}^{2}x_{B}^{2}}{Q^{2}}.
\end{equation}

The amplitude of the axial current transition into an on-shell pion
in the diagram Figure~\ref{fig:DVMPLT}~ (c) according to PCAC has
the form, 
\begin{equation}
\left\langle 0\left|J_{\mu}^{b,5}(0)\right|\pi^{a}(q-\Delta)\right\rangle =if_{\pi}\sqrt{2}q_{\mu}
\end{equation}
In order to simplify the calculation of the leptonic part of the diagram
(c), we employ the chain of identities 
\begin{equation}
\hat{k}_{\pi}\left(1-\gamma_{5}\right)\nu\left(k_{\nu}\right)=\left(1+\gamma_{5}\right)\hat{k}_{\pi}\nu\left(k_{\nu}\right)=\left(1+\gamma_{5}\right)\left(\hat{k}_{\pi}-\hat{k}_{\nu}\right)\nu\left(k_{\nu}\right),\label{eq:Aux_1}
\end{equation}
\begin{align}
S\left(k_{\nu}-k_{\pi}\right)\left(1+\gamma_{5}\right)\left(\hat{k}_{\pi}-\hat{k}_{\nu}\right) & =S\left(k_{\nu}-k_{\pi}\right)\left(\hat{k}_{\pi}-\hat{k}_{\nu}\right)\left(1-\gamma_{5}\right)\label{eq:Aux_2}\\
 & =-1+\gamma_{5}+m_{\mu}S\left(k_{\nu}-k_{\pi}\right)\approx-\left(1-\gamma_{5}\right),\nonumber 
\end{align}

\begin{align}
\mathcal{A}_{l} & \sim e\bar{\mu}\left(k_{\mu}\right)\hat{\epsilon}\left(k_{\gamma}\right)S\left(k_{\nu}-k_{\pi}\right)\hat{k}_{\pi}\left(1-\gamma_{5}\right)\nu\left(k_{\nu}\right)\label{eq:Ampl_def}\\
 & =-e\bar{\mu}\left(k_{\mu}\right)\hat{\epsilon}\left(k_{\gamma}\right)\nu\left(k_{\nu}\right)+\mathcal{O}\left(m_{\mu}\right),\nonumber 
\end{align}
where in~(\ref{eq:Aux_1}) we make use of the fact that the initial
state neutrino is on-shell, $\hat{k}_{\nu}\nu(k)=0$. Actually, the
simplification of~(\ref{eq:Ampl_def}) is a manifestation of the
Ward-Takahashi-Slavnov-Taylor identity for the charged lepton current,
\begin{align}
i\partial_{\alpha}\left(\bar{\mu}\gamma_{\alpha}\left(1-\gamma_{5}\right)\nu\right) & =-\frac{g}{\sqrt{2}}\bar{\nu}\hat{W}^{+}\left(1-\gamma_{5}\right)\nu+\frac{g}{\sqrt{2}}\bar{\mu}\hat{W}^{+}\left(1-\gamma_{5}\right)\mu-e\bar{\mu}\hat{A}\left(1-\gamma_{5}\right)\nu\nonumber \\
 & -\frac{g\left(g_{V}^{(\mu)}+g_{A}^{(\mu)}-g_{V}^{(\nu)}-g_{A}^{(\nu)}\right)}{2\cos\theta_{W}}\bar{\mu}\hat{Z}\left(1-\gamma_{5}\right)\nu\not=0.\label{eq:WardTakahashi}
\end{align}
This simplification explains why there is no harmonics in the denominator
of the BH and interference terms, and it is valid in the limit of
massless leptons. The diagrams (b, c) yield for the amplitude (sign
corresponds to $\pi^{+}$) 
\begin{equation}
\sim e^{2}\frac{\bar{\mu}(k-q)\gamma_{\mu}(1-\gamma_{5})\nu(k)}{t}if_{\pi}\left(-g_{\mu\nu}+\frac{\mathcal{A}_{\mu\nu}}{1+Q^{2}/M_{W}^{2}}\right)\bar{U}\left(P+\Delta\right)\left(F_{1}(t)\gamma_{\nu}+\frac{i\sigma_{\nu\alpha}\Delta_{\alpha}}{2M}F_{2}(t)\right)U\left(P\right)\label{eq:BH_Amp-1}
\end{equation}

Further evaluation of Bethe-Heitler~(\ref{eq:XSec_BH}) and interference~(\ref{eq:XSec_Int})
terms requires some trivial but tedious Dirac algebra, which was done
with \textsc{FeynCalc}~\cite{Mertig:1990an}.

\section*{Acknowledgments}

This work was supported in part by Fondecyt (Chile) grants No. 1090291,
1100287 and 1120920.

 \end{document}